\documentstyle[aps,prb,epsf]{revtex}

\newcommand{\tg}{\tilde{g}}

\begin{document}
\draft
\twocolumn[\hsize\textwidth\columnwidth\hsize\csname
@twocolumnfalse\endcsname

\title{Dynamics of a Josephson Array in a Resonant Cavity}

\author{E. Almaas\cite{email1} and D. Stroud\cite{email2}}
\address{Department of Physics, The Ohio State University, Columbus,
Ohio 43210}
\date{\today}
\maketitle
\begin{abstract}
We derive dynamical equations for a Josephson array coupled to a
resonant cavity by applying the Heisenberg equations of motion to a
model Hamiltonian described by us earlier \protect{[}Phys. Rev. B {\bf
63}, 144522 (2001); Phys. Rev. B {\bf 64}, 179902 (E)\protect{]}.  By
means of a canonical transformation, we also show that, in the absence
of an applied current and dissipation, our model reduces to one
described by Shnirman {\it et al} \protect{[}Phys.\ Rev.\ Lett.\ {\bf
79}, 2371 (1997)\protect{]} for coupled qubits, and that it
corresponds to a capacitive coupling between the array and the cavity
mode.  From extensive numerical solutions of the model in one
dimension, we find that the array locks into a coherent, periodic
state above a critical number of active junctions, that the
current-voltage characteristics of the array have self-induced
resonant steps (SIRS's), that when $N_a$ active junctions are
synchronized on a SIRS, the energy emitted into the resonant cavity is
quadratic in $N_a$, and that when a fixed number of junctions is
biased on a SIRS, the energy is linear in the input power.  All these
results are in agreement with recent experiments.  By choosing the
initial conditions carefully, we can drive the array into any of a
variety of different integer SIRS's.  We tentatively identify terms in
the equations of motion which give rise to both the SIRS's and the
coherence threshold.  We also find higher-order integer SIRS's and
fractional SIRS's in some simulations.  We conclude that a resonant
cavity can produce threshold behavior and SIRS's even in a
one-dimensional array with appropriate experimental parameters, and
that the experimental data, including the coherent emission, can be
understood from classical equations of motion.
\end{abstract}

\pacs{PACS numbers: 05.45.Xt, 79.50.+r, 05.45.-a, 74.40.+k}
\vskip1.5pc] 

\section{Introduction.}

A long-standing goal of experimental \cite{booi,han,kaplunenko,wan}
and theoreti-cal
\cite{hadley,octavio,wiesenfeld1,braiman,darula,whan,filatrella1,filatrella}
research on Josephson junction arrays has been to develop sources of
coherent microwave radiation.  The basic idea underlying this work is
that a Josephson junction is a simple way of converting a d.\ c.
current into an a.\ c. voltage.  Thus, an array of $N$ Josephson
junctions oscillating in phase should produce a signal with $N$ times
the voltage amplitude, and hence $N^2$ times the emitted a.\ c.
power, of a single junction.  Arrays of overdamped junctions have
seemed most promising for coherent emission, since junctions of this
type have, at any given applied current, only a single voltage state,
and thus have none of the multistability and chaotic behavior which
could inhibit coherent emission. However in practice, it has proved
very difficult to achieve an efficient a.\ c. to d.\ c. conversion in
such systems; thus far the highest conversion efficiency is only about
1\% \cite{jain,benz,cawthorne1}.  The low efficiency may result from
the high degree of neutral stability which has been shown to exist in
such overdamped arrays in the absence of an applied magnetic field
\cite{neutral}.

Recently, a remarkably high conversion efficiency has been achieved in
an {\em underdamped} Josephson array, by coupling the junctions in
that array electromagnetically to a mode in a resonant microwave
cavity \cite{barbara}.  The high emission is a manifestation of the
so-called self-induced resonant steps (SIRS's) that appear on the
current-voltage (IV) characteristics of these arrays.  It is thought
such arrays emit strongly because every junction is coupled to the
same electromagnetic mode, and hence, effectively to every other
junction.  This same coupling is presumed to lead to the observed
threshold effect, in which the strong emission occurs only above a
certain array length.
 
A number of models have been proposed which produce some aspects of
this behavior.  In the original model, which inspired the
measurements, an analogy was drawn between junctions in a
voltage-biased series Josephson array and a collection of two-level
atoms where population inversion and laser emission could be achieved
\cite{bonifacio}.  Several authors have introduced various kinds of
impedance loads across groups of junctions or a one-dimensional array,
in order to achieve a global coupling and hence, to investigate
coherence among the junctions in the array
\cite{filatrella,wiesenfeld,cawthorne,abdullaev}.  None of these
models have yet produced both the self-induced resonant steps and the
threshold junction number seen in the experiments.  The coupling of
propagating modes in Josephson ladders and other structures to
electromagnetic radiation has also been studied theoretically
\cite{fistul,jensen}.  A simple Hamiltonian to treat the equilibrium
properties of a one-dimensional, voltage-driven array in the
weak-coupling regime has recently been proposed and studied within a
mean-field approximation which should be very accurate in the limit of
large numbers of junctions \cite{harb}.  Within this approach, it was
found that the array developed coherence only above a threshold number
of junctions, in agreement with experiment.  In a recent paper
\cite{almaas}, the present authors proposed a similar model to treat
array {\em dynamics} for any strength of coupling; they also briefly
described a few numerical results obtained from the model, including
both a thresh-
\begin{figure}[t]
\epsfysize=5cm \centerline{\epsffile{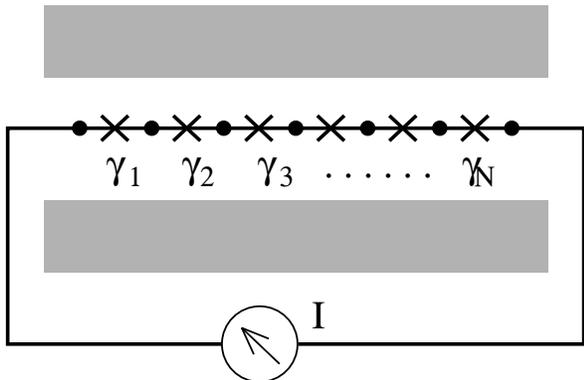}}
\caption{Sketch of the array geometry considered in our model.  There
         are $N$ junctions (crosses), labeled by their gauge-invariant
         phase differences $\gamma_j$, and $N + 1$ superconducting
         islands.  A current $I$ is injected into one end of the array
         and extracted from the other.  The array is placed in an
         electromagnetic cavity which supports a single resonant
         photon mode of frequency $\Omega$.}
\label{fig:geometry}
\end{figure}
\noindent old for coherence and self-induced resonant steps.

In the present paper, we give a more complete derivation of the model
equations of motion of Ref.\ [25] for a one-dimensional array of
underdamped Josephson junctions coupled to a single-mode
electromagnetic cavity. Starting from a suitable Hamiltonian, we
obtain the Heisenberg equations of motions for the phase differences
and the photon creation and annihilation operators.  We account for
dissipation in the junctions by the standard procedure of coupling
each junction to a reservoir of phonon variables with a density of
states constructed so as to produce ohmic damping.  In the limit of
large numbers of photons, the equations can be treated classically and
solved numerically.  We also correct the treatment of Ref.\ [25] of
the junction damping and the coupling of the array to an external
current.  Finally, we carry out a canonical transformation of our
Hamiltonian to show that the interaction between the array and the
cavity mode has the form of a {\em capacitive} coupling.

We also present much more extensive numerical results than those of
Ref.\ [25], based on solutions to the model equations.  Our numerical
results show all the principal features of the measurements, including
SIRS's, a coherence threshold, and a quadratic dependence of the
photon energy in the cavity upon number of active junctions.  Plots of
the IV characteristics and other calculated features closely resemble
the corresponding experimental plots.  This agreement is especially
noticeable since the calculation is one-dimensional, while most
experiments have been conducted on two-dimensional arrays.  In
addition, we find that by tuning the initial conditions, we can cause
the array to lock into a variety of different integer SIRS's, again in
agreement with experiment.  We conclude that these equations do indeed
describe the experiment, and that a one-dimensional array is
sufficient to achieve this type of coherent behavior.

The remainder of this paper is organized as follows.  In Section II,
we derive the Heisenberg equations of motion for the phase and photon
variables, starting from a model Hamiltonian.  We also apply a
canonical transformation which shows that the Hamiltonian involves a
kind of distributed capacitive coupling between the Josephson array
and the cavity mode.  In Section III, we give a detailed description
of our numerical results.  Section IV presents a comparison between
our results and experiment, gives a qualitative discussion of the
numerical results, and makes some concluding remarks about the model.

\section{Derivation of the Equations of Motion}

\subsection{Hamiltonian}

We consider a one-dimensional array of N Josephson junctions placed in
a resonant cavity, which we assume supports only a single photon mode
of frequency $\Omega$ (the geometry is sketched in Fig.\
\ref{fig:geometry}).  The array is to be driven by an applied current
$I$.  We write the Hamiltonian in the form
\begin{equation}
H = H_{photon} + H_J + H_C + H_{curr} + H_{diss}.
\label{eq:ham}
\end{equation}
Here, $H_{photon}$ is the energy of the cavity mode, which we express
as
\begin{equation}
H_{photon} = \hbar\Omega\left(a^\dag a + \frac{1}{2}\right),
\end{equation}
with $a^\dag$ and $a$ as the usual photon creation and annihilation
operators.  $H_J$ is the Josephson Hamiltonian, and is assumed to take
the form
\begin{equation}
H_J = -\sum_{j=1}^N E_{Jj}\cos\gamma_j,
\end{equation}
where $E_{Jj}$ is the Josephson energy of the j$^{th}$ junction, and
$\gamma_j$ is the gauge-invariant phase difference across the j$^{th}$
junction (defined more precisely below).  $E_{Jj}$ is related to
$I_{cj}$, the critical current of the j$^{th}$ junction, by $E_{Jj} =
\hbar I_{cj} /q$, where $q = 2|e|$ is the Cooper pair charge.  $H_C$
is the capacitive energy of the N junctions, which we approximate as
\begin{equation}
H_C = \sum_{j=1}^N E_{Cj}n_j^2.
\end{equation}
Here, $E_{Cj} = q^2/(2C_j)$ is the capacitive energy of the j$^{th}$
junction, $C_j$ is the capacitance of that junction, and $n_j$ is the
difference between the number of Cooper pairs on the j$^{th}$ and
$(\mbox{j}+1)^{th}$ superconducting island.

The gauge-invariant phase difference, $\gamma_j$, is the term which
couples the Josephson junctions to the cavity.  It may be written as
\begin{equation}
\gamma_j = \phi_j - [(2\pi)/\Phi_0] \int_j
	{\bf A} \cdot {\bf ds} \equiv~ \phi_j - A_j, \label{eq:gauge}
\end{equation} 
where $\phi_j$ is the phase difference across the j$^{th}$ junction in
a particular gauge, ${\bf A}$ is the vector potential in that same
gauge, $\Phi_0 = hc/q$ is the flux quantum, and the line integral is
taken across the junction.  We assume that ${\bf A}$ arises from the
electromagnetic field of the normal mode of the cavity.  In Gaussian
units, this vector potential takes the form\cite{slater,yariv}
\begin{equation}
{\bf A}({\bf x},t) = \sqrt{(h c^2) / (\Omega)} \left(a(t) +
       a^\dag(t)\right){\bf E}({\bf x}),
\end{equation}
where ${\bf E}({\bf x})$ is a vector proportional to the local
electric field of the mode, normalized such that $\int_Vd^3x|{\bf
E}({\bf x})|^2 = 1$, where $V$ is the cavity volume.  Given this
representation for ${\bf A}$, the phase factor $A_j$ can be written
\begin{equation}
A_j = \sqrt{g_j}(a + a^\dag),
\end{equation}
where $g_j$ takes the form
\begin{equation}
g_j = \frac{\hbar c^2}{\Omega} \frac{(2\pi)^{3}} {\Phi_0^2}
\left[\int_j{\bf E}({\bf x})\cdot{\bf ds}\right]^2.
\end{equation}
Clearly, $g_j$ is an effective coupling constant describing the
interaction between the j$^{th}$ junction and the cavity\cite{gauge}.

The terms discussed so far need to be supplemented by the effects of a
driving current and of damping within the junctions.  A driving
current is easily included within the Hamiltonian formalism via a
``washboard potential,'' H$_{curr}$, of the form
\begin{equation}
H_{curr} = -\frac{\hbar I}{q}\sum_{i=1}^N\gamma_j,
\label{eq:wash}
\end{equation}
with $I$ as the driving current.

The inclusion of dissipation can be done in a standard way
\cite{chakra,leggett,ambeg} by coupling each gauge-invariant phase
difference, $\gamma_j$, to a separate collection of harmonic
oscillators with a suitable spectral density.  Thus, we write the
dissipative term in the Hamiltonian as
\begin{equation}
H_{diss} = \sum_{j=1}^N H_{diss, j},
\end{equation}
where
\begin{eqnarray}
H_{diss,j} &=& \sum_{\alpha} \left[f_{\alpha,j}~ \gamma_j u_{\alpha,j} +
	\frac{p_{\alpha,j}^2}{2m_{\alpha,j}} \nonumber \right.\\
         &+& \left.  \frac{1}{2}~ m_{\alpha,j}~
	\omega_{\alpha,j}^2 ~u_{\alpha,j}^2 + \frac{(f_{\alpha,j})^2}{
	2 m_{\alpha,j} ~\omega_{\alpha,j}^2} (\gamma_j)^2  \right].
	\label{eq:diss}
\end{eqnarray}
The variables $u_{\alpha,j}$ and $p_{\alpha,j}$, describing the
$\alpha^{th}$ oscillator in the j$^{th}$ junction, are canonically
conjugate, and $m_{\alpha,j}$ and $\omega_{\alpha,j}$ are the mass and
frequency of that oscillator.  The last term in Eq. (\ref{eq:diss})
must be added in order to prevent the original potential from being
shifted by the coupling to the j$^{th}$ phase degree of freedom
\cite{leggett}. The spectral density of the harmonic oscillators in
the j$^{th}$ junction, denoted $J_j(\omega)$, is defined by
\begin{equation}
J_j(\omega) \equiv ( \frac{\pi}{2}) \sum_\alpha 
   \frac{(f_{\alpha,j})^2} {m_\alpha \omega_\alpha}~ \delta ( \omega -
   \omega_\alpha ).
\label{eq:spec}
\end{equation}
If $J_j(\omega)$ is linear in $|\omega|$, it can be shown that the
dissipation in the junction is ohmic \cite{chakra,leggett,ambeg}.  We
write such a linear spectral density as
\begin{equation}
J_j(\omega) = \frac{\hbar}{2\pi}~ \alpha_j~ |\omega|~ \Theta (\omega_c -
              \omega ),
\end{equation}
where $\omega_c$ is a high-frequency cutoff (at which the assumption
of ohmic dissipation begins to break down), $\Theta(\omega_c -
\omega)$ is the usual step function, and $\alpha_j$ is a dimensionless
constant, which we write as $\alpha_j = R_0/R_j$, where $R_0 =
h/(4e^2)$ and $R_j$ is a constant with dimensions of resistance
(actually, the effective shunt resistance of the junction, as
discussed below).

\subsection{Equations of Motion}

It is now convenient to introduce the operators $a_R$ and $a_I$ by
\begin{equation}
a = a_R + ia_I;
\end{equation}
\begin{equation}
a^\dag = a_R - ia_I.
\end{equation}
The free photon part of the Hamiltonian can be expressed in terms of
$a_R$ and $a_I$ as follows:
\begin{equation}
H_{photon} = \hbar\Omega(a_R^2 + a_I^2),
\label{eq:ephot}
\end{equation}
where we have used the additional commutation relations $[a_R, a_I] =
i/2$, which follows from the usual relation $[a, a^\dag] = 1$.  The
gauge-invariant phase difference, $\gamma_j$, is related to $\phi_j$
by
\begin{equation}
\gamma_j = \phi_j - 2\sqrt{g_j}a_R.
\end{equation}

The time-dependence of the various operators appearing in the
Hamiltonian (\ref{eq:ham}) is now obtained from the Heisenberg
equations of motion.  For a general operator ${\cal O}$, these take
the form
\begin{equation}
\dot{{\cal O}} = \frac{1}{i\hbar}[{\cal O}, H].
\end{equation}
These equations of motion can be evaluated for the various operators
entering $H$, using the commutation relation $[A, F(B)] = [A,
B]F^\prime(B)$, where $F$ is any function of an operator B, and
$F^\prime(B)$ is the derivative of that function.  One also needs the
commutation relations for the various operators in the Hamiltonian
(\ref{eq:ham}).  Besides the relations already given, these are as
follows:
\begin{eqnarray}
\protect{[}n_j, \gamma_k\protect{]} &=& -i \delta_{jk};\\
\protect{[}p_{\alpha,j}, u_{\beta,k}\protect{]} &=& 
    -i\hbar~ \delta_{\alpha,\beta}~ \delta_{j,k}.
\end{eqnarray}
Note that $\gamma_k$, unlike $\phi_j$, no longer commutes with $a_I$;
instead, it satisfies 
\begin{eqnarray}
\protect{[} \gamma_j, a_R \protect{]} &=& 0; \\
\protect{[} \gamma_j, a_I \protect{]} &=& -i \sqrt{g_j}.
\end{eqnarray}

Using all these relations, we find, after a little algebra, the
following equations of motion for the operators $\gamma_j$, $n_j$,
$a_R$, and $a_I$:
\begin{eqnarray}
\label{eq:eom1}
\dot{\gamma}_j &= &2~ \frac{E_{Cj}}{\hbar}~ n_j - 2~ \Omega~
        \sqrt{g_j}~ a_I,\\
\label{eq:eom2}
\dot{n}_j & = &-\frac{E_{Jj}}{\hbar} \sin(\gamma_j) + \frac{I}{q}
	\nonumber \\
        &&-\frac{1}{\hbar}\sum_{\alpha} \left( f_{\alpha,j}~ 
	u_{\alpha,j} + \frac{(f_{\alpha,j})^2}{m_{\alpha,j} 
	\omega_{\alpha,j}^2}~ \gamma_j \right),\\
\label{eq:eom3}
\dot{a}_R &=& \Omega~ a_I,\\
\label{eq:eom4}
\dot{a}_I &=&-\Omega~ a_R + \sum_j \sqrt{g_j}~ \frac{E_{Jj}}{\hbar}~ 
          \sin(\gamma_j) - \frac{I}{q} \sum_j \sqrt{g_j} \nonumber\\
      && +\sum_j \frac{\sqrt{g_j}}{\hbar}~ \sum_{\alpha} 
          \left(f_{\alpha,j}~ u_{\alpha,j} + \frac{(f_{\alpha,j})^2}
	{m_{\alpha,j} \omega_{\alpha,j}^2}~ \gamma_j \right).
\end{eqnarray}
These are equations of motion for the {\em operators} $a_R$, $a_I$,
$n_j$, and $\phi_j$ (or $\gamma_j$).  Note that they do not depend on
the particular choice of gauge, but only on the form of the
Hamiltonian and the commutation relations for the various operators.
We will study these general equations within the limit of large number
of photons in the cavity and large number of charges in the junctions,
and in this ``classical'' limit, we will regard the operators as
$c$-numbers \cite{classical}.

We also have the equations of motion for the harmonic oscillator
variables. Since we have no explicit interest in these variables for
themselves, we instead eliminate them in order to incorporate the
dissipative term into the equations of motion.  Such a replacement is
possible provided that the spectral density of each junction is linear
in frequency, as noted above.  In that case
\cite{chakra,leggett,ambeg}, the oscillator variables can be
integrated out.  The effect of carrying out this procedure is that one
should make the replacement
\begin{equation}
\sum_{\alpha} \left( f_{\alpha,j}~ u_{\alpha,j} +  \frac{(f_{\alpha,j})^2}
	{m_{\alpha,j} \omega_{\alpha,j}^2}~ \gamma_j \right) \rightarrow
       \frac{\hbar}{2\pi}\frac{R_0}{R_j}~ \dot{\gamma}_j
\label{eq:subst}
\end{equation}
wherever this sum appears in the equations of motion.

Making the replacement (\ref{eq:subst}) in Eqs.\ (\ref{eq:eom2}) and
(\ref{eq:eom4}), we obtain the equations of motion for $n_j$ and $a_I$
with damping:
\begin{eqnarray}
\label{eq:eom2a}
\dot{n}_j &=& -\frac{E_{Jj}}{\hbar} \sin(\gamma_j) + \frac{I}{q}
            -\frac{\bar{\omega}_p}{2 \omega_{Cj} Q_{Jj}}~ \dot{\gamma}_j\\
\label{eq:eom4a}
\dot{a}_I &=& -\Omega~ a_R + \sum_j \sqrt{g_j}~ \frac{E_{Jj}}{\hbar}~ 
        \sin(\gamma_j) \nonumber \\
      && - \frac{I}{q} \sum_j \sqrt{g_j} + \sum_j \sqrt{g_j}~
        \frac{\bar{\omega}_p}{2 \omega_{Cj} Q_{Jj}}~ \dot{\gamma_j}.
\end{eqnarray}
Here, we have introduced the parameters $\omega_{Cj} = E_{Cj}/\hbar$,
which is a frequency associated with the capacitive energy of the
j$^{th}$ junction; $\bar{\omega}_p = \frac{1}{N}\sum_{j =
1}^N\omega_{pj}$, the average of the Josephson plasma frequencies
$\omega_{pj}=\sqrt{2E_{Cj} E_{Jj}}/\hbar$; and $Q_{Jj}$, the
dimensionless junction quality factor (or damping parameter) for the
j$^{th}$ junction, which is related to the capacitance $C_j$ and the
shunt resistance $R_j$ by
\begin{equation}
Q_{Jj}=\bar{\omega}_p R_jC_j.
\end{equation}

Eqs.\ (\ref{eq:eom1}), (\ref{eq:eom3}), (\ref{eq:eom2a}), and
(\ref{eq:eom4a}) can be combined, with a little algebra, into two
coupled second-order differential equations:
\begin{eqnarray}
\frac{1}{2\omega_{Cj}}~ \ddot{\gamma}_j + \frac{\bar{\omega}_p}{2
    \omega_{Cj}Q_{Jj}}~ \dot{\gamma}_j &+& \omega_{Jj}~\sin\gamma_j
	\nonumber \\
    &=& \frac{I}{q} - \frac{\sqrt{g_j}}{\omega_{Cj}} ~\ddot{a}_R
\label{eq:eom1b}
\end{eqnarray}
and
\begin{equation}
\left(1 + \Omega \sum_j \frac{g_j}{\omega_{Cj}} \right) \ddot{a}_R +
    \Omega^2~ a_R = -\frac{\Omega}{2}~ \sum_j
    \frac{\sqrt{g_j}}{\omega_{Cj}} ~\ddot{\gamma}_j.
\label{eq:eom2b}
\end{equation}
where we have defined $\omega_{Jj} = E_{Jj}/\hbar$. Note that in the
absence of coupling between the junctions and the cavity, $\gamma_j =
\phi_j$, Eq.\ (\ref{eq:eom1b}) reduces to the usual equation of motion
for a resistively and capacitively shunted junction\cite{tinkham}
driven by a current $I$, and Eq.\ (\ref{eq:eom2b}) reduces to that of
a simple harmonic oscillator which represents the cavity mode.  Note
that we have not included any damping due to the cavity walls.  While
such damping is undoubtedly present, we find that good agreement with
experiment can be obtained without including it.

\subsection{Canonical Transformation}

The physics behind the coupling between the Josephson junctions and
the resonant cavity, and hence the physics of Eqs.\ (\ref{eq:eom1}) -
(\ref{eq:eom4}), can be made clearer by a canonical transformation.
For simplicity, we describe this transformation including only the
terms H$_{phot}$, H$_J$, and $H_C$ from the Hamiltonian
(\ref{eq:ham}), and omitting $H_{curr}$ and $H_{diss}$.  The same
transformation has previously been used for a {\em single} junction
coupled to a resonant cavity by Buisson and Hekking\cite{buisson};
and, for two voltage-driven junctions coupled to a resonant cavity, by
Shnirman {\it et al} \cite{shnirman}.

We begin by writing
\begin{eqnarray}
H^\prime &\equiv& H_{phot} + H_J + H_C ~=~\frac{1}{2} \hbar~ \Omega~ 
	( p_r^2 + q_r^2 ) \nonumber \\
      &&~+~ \sum_{j=1}^N \bigg[  E_{Cj} n_j^2 - E_{Jj} 
	\cos (\phi_j - \sqrt{2 g_j}~ q_r) \bigg], \label{eq:ham2}
\end{eqnarray}
where we have defined $p_r = \sqrt{2}~ a_I$ and $q_r = \sqrt{2}~ a_R$.
With this choice, $p_r$ and $q_r$ satisfy the commutation relation
$[p_r, q_r] = -i$.  Next, we make the canonical transformation
\begin{eqnarray}
n^\prime_j    &=& n_j; \\
\phi^\prime_j  &=& \phi_j - \sqrt{2 g_j}~ q_r; \\
p^\prime_r    &=& p_r + \sum_{j=1}^N \sqrt{2 g_j}~ n_j; \\
q^\prime_r    &=& q_r.
\end{eqnarray}
The only nonvanishing commutators of the primed variables are easily
shown to be $[n^\prime_j,\phi^\prime_j] = [p^\prime_r,q^\prime_r] =
-i$. Re-expressing the Hamiltonian (\ref{eq:ham2}) in terms of the
primed variables, we obtain
\begin{eqnarray}
H^\prime &=& \frac{1}{2} \hbar \Omega 
\left (p^\prime_r - \sum_j\sqrt{2g_j}n_j^\prime\right)^2 +
   \frac{1}{2}\hbar\Omega(q^\prime_r)^2  \nonumber \\
	&& +~ \sum_{j=1}^N \bigg[ E_{Cj} (n^\prime_j)^2 - E_{Jj} 
	\cos \phi^\prime_j\bigg] \label{eq:canon}
\end{eqnarray}
Thus, $H^\prime$ is the sum of four terms: the sum $(\hbar \Omega / 2)
[(p_r^\prime)^2 + (q_r^\prime)^2]$ describes the cavity resonator; the
last sum describes the $N$ independent junctions; and the remaining
terms represent the interaction between the junctions and the
resonator, and an indirect interaction between the junction variables
$n_j^\prime$ mediated by the cavity.

To interpret this interaction, we note that the junction-cavity system
has two places in which charge can be stored: the variables
$n_j^\prime$ of the junctions and the variables $p_r^\prime$ of the
cavity.  The cavity behaves much like an $LC$ circuit, with capacitive
energy $(\hbar \Omega / 2) (p_r^\prime)^2$ and inductive energy
$(\hbar \Omega / 2) (q_r^\prime)^2$.  The dominant interaction is a
{\em capacitive} coupling between the charge variables $n_j^\prime$ of
the junction and the charge variable $p_r^\prime$ of the cavity\cite{asymm}.

In further support of this interpretation, we now show that
$H^\prime$, in the form (\ref{eq:canon}), is equivalent to that given
by Shnirman {\it et al} \cite{shnirman} in the case of zero applied
voltage.  Fig.\ 1 of Ref.\ [35] depicts each junction contained
between the plates of a capacitor with capacitance $C$.  The junction
itself has capacitance $C_J$.  The system of junction and capacitor is
then shunted by a parallel inductance, $L$, which acts to couple all
the junctions together \cite{note2}. The equivalence is established by
noting the correspondence between the variables used in the present
paper and the variables $\phi$, $q$, $n_j$ and $\theta_j$ of Ref.\
[35].  The correspondence is as follows (assuming that the coupling
constants $g_j$, $E_{Cj}$, and $E_{Jj}$ are independent of $j$):
\begin{eqnarray}
q_r^\prime    &\leftrightarrow& [L/(2C_t)]^{-1/4}\phi; \nonumber \\
p_r^\prime    &\leftrightarrow& [L/(2C_t)]^{1/4}q; \nonumber \\
n_j^\prime    &\leftrightarrow& n_j \nonumber \\
\phi_j^\prime &\leftrightarrow& \theta_j, \nonumber
\end{eqnarray}
where 
$C_t = C C_J / (C + 2 C_J)$.  In order to complete the correspondence, we
also give the correspondence between the remaining parameters 
and the quantities $L$, $C$, $C_t$,and $C_J$:
\begin{eqnarray}
\sqrt{2g} &=& \left(\frac{L}{2C_t}\right)^{1/4} \frac{C_t}{C_J};
		\nonumber \\ 
\Omega    &=& \frac{1}{\sqrt{2LC_t}}; \nonumber \\ 
E_C       &=& \frac{1}{C + 2C_J}, \nonumber
\end{eqnarray}
If we make the replacements and identifications given above, then our
Hamiltonian, for the case of two junctions, is identical to that
considered in [35] in the absence of an applied voltage.  The main
differences between the two models are the boundary conditions:
constant current bias in our model, and fixed voltage bias in that of
Ref.\ [35].

\subsection{Dimensionless Form}

In order to write these equations of motion in a simpler form, we now
introduce the dimensionless time $\tau=\bar{\omega}_pt$.  Also,
although we allow disorder in the parameters of the different
Josephson junctions, we assume that the coupling constants between the
junctions and the cavity are all the same, i. e., that $g_j = g$ for
all $j$.  In addition, for simplicity, we assume that the products
$R_jC_j$ and $I_{cj}/C_j$, and hence $\omega_{pj}$, are independent of
$j$.  Then $Q_{Jj}$ is also $j$-independent, and we may introduce the
dimensionless coupling constant
\begin{equation}
\tilde{g}= g ~\frac{E_{J}}{\hbar\bar{\omega}_p},
\end{equation}
which is also independent of $j$.  Similarly, we introduce a scaled
number variable $\tilde{n}_j$ by
\begin{equation}
\tilde{n}_j= \frac{2 E_{Cj}}{\hbar\bar{\omega}_p}~ n_j,
\end{equation} 
a dimensionless frequency $\tilde{\Omega}$ by $\tilde{\Omega} =
\Omega/ \bar{\omega}_p$, and scaled photon variables $\tilde{a}_R$ and
$\tilde{a}_I$ by
\begin{eqnarray}
\tilde{a}_R & = & \sqrt{g}~ a_R;\\
\tilde{a}_I & = & \sqrt{g}~ a_I.
\end{eqnarray}
Finally, we assume that the critical currents $I_{cj}$ are random and
uniformly distributed between $I_c(1-\Delta)$ and 
\begin{figure}[t]
\epsfysize=6.6cm
\centerline{\epsffile{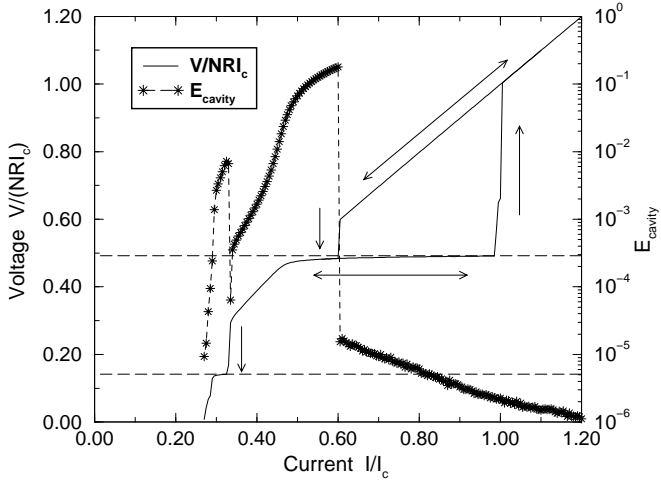}}
\caption{Left-hand scale and solid line: current-voltage (IV)
         characteristics of a one dimensional array of $N=40$
         junctions, with disorder parameter $\Delta = 0.05$ and
         coupling constant $\tg = 0.001$.  The resonant frequency of
         the cavity is $\tilde{\Omega} = 2.2$, and the damping
         parameter of the junctions is $Q_J = \sqrt{20}$.  Right-hand
         scale and stars : scaled total energy $\tilde{E} = g
         E/(\hbar\Omega)$ carried by the resonant mode of the cavity,
         plotted as a function of decreasing current $I/I_c$.  The
         vertical dashed lines are guides to the eye.  The upper
         dashed horizontal line indicate the expected position of the
         integer self-induced resonant steps (SIRS's) for the
         particular resonant frequency $\tilde{\Omega}$ of the cavity
         (all junctions in the $n=1$ SIRS). For the lower dashed
         horizontal line, 23 junctions are on the $n=1/2$ SIRS, and 17
         junctions are in the $\langle V_j \rangle_\tau = 0$
         state. Branches corresponding to increasing and decreasing
         current are shown by arrows.  Double-headed arrows on this
         and subsequent Figures denote that the curve can be obtained
         by sweeping the current in either direction.}
\label{fig:IV_cavity}
\end{figure}
\noindent
$I_c(1+\Delta)$, where $\Delta$ is a measure of the degree of
disorder.  After some algebra, we eventually obtain the following
equations of motion:
\begin{eqnarray}
\dot{\gamma}_j    &=& \tilde{n}_j -2~ \tilde{\Omega}~ \tilde{a}_I; \\
\dot{\tilde{n}}_j &=& \frac{I}{I_c (1+\Delta_j)} -\frac{\tilde{n}_j}{Q_J}
               - \sin(\gamma_j) + 2~\frac{\tilde{\Omega}}{Q_J} 
                   ~\tilde{a}_I; \label{eq:1}\\
\dot{\tilde{a}}_R &=& \tilde{\Omega}~ \tilde{a}_I; \\
\dot{\tilde{a}}_I &=&-\tilde{\Omega}~ \tilde{a}_R - 2~ \tilde{\Omega}~ 
	        \tilde{g}~ \frac{\tilde{a}_I}{Q_J}~ \sum_j (1+\Delta_j)
	        - N~ \tilde{g}~ \frac{I}{I_c} \nonumber \\
              &+&\tilde{g}~\sum_j (1+\Delta_j) \sin(\gamma_j) + 
	      \frac{\tilde{g}}{Q_J} ~\sum_j (1+\Delta_j)~ \tilde{n}_j. 
		\label{eq:2}
\end{eqnarray}
In these equations, the dot refers to differentiation with respect to
$\tau$, and the j$^{th}$ critical current is $I_c(1+\Delta_j)$.

These equations can be combined into two more compact equations, with
the result
\begin{eqnarray}
\ddot{\gamma}_j + \frac{1}{Q_J}~ \dot{\gamma}_j + \sin (\gamma_j) 
        & = & \frac{I}{I_c (1+\Delta_j)} - 2~ \ddot{\tilde{a}}_R ;
        \label{eq:ddgam1}\\
\ddot{\tilde{a}}_R + (\Omega')^2~ \tilde{a}_R &=& - \tg~ \frac{\Omega'}
	{\tilde{\Omega}} ~\sum_{j=1}^N (1+\Delta_j)~ \ddot{\gamma}_j,
        \label{eq:ddgam2}
\end{eqnarray}
where we have defined $(\Omega')^2 ~=~ \tilde{\Omega}^2/[1+2~ \tg~
\tilde{\Omega} \sum_j (1+\Delta_j)]$.  Eqs.\ (\ref{eq:ddgam1}) and
(\ref{eq:ddgam2}) are the analogs of Eqs.\ (\ref{eq:eom1b}) and 
(\ref{eq:eom2b}), expressed in terms of dimensionless reduced variables.

\section{Numerical Results}

We have solved Eqs. (\ref{eq:ddgam1}) and (\ref{eq:ddgam2}) for the
variables $\tilde{n}_j$, $\gamma_j$, $\tilde{a}_R$ and $\tilde{a}_I$
numerically using the same approach as in Ref.\ [25], namely, by
implementing the rapid and accurate adaptive Bulrisch-Stoer method
\cite{numrec}.  We initialize the simulations with all the phases
randomized between $[0,2\pi]$, and usually $\tilde{a}_R = \tilde{a}_I
= \tilde{n}_j = 0$. We then let the system equilibrate for a time
interval $\Delta\tau=10^4$, after which we evaluate averages over a
time interval $\Delta\tau=2\cdot 10^3$, using $2^{16}$ evenly spaced
sampling points.

\subsection{Typical IV Characteristics, Power Spectrum, and Coherence
Transition}

In Fig.\ \ref{fig:IV_cavity}, we show a representative current-voltage
(IV) characteristic calculated for an array of $N = 40$ junctions with
$\Delta=0.05$ and $\tg = 0.001$.  The time-averaged voltage $\langle V
\rangle_\tau$ [left-hand scale] is obtained from
\begin{equation}
\langle V \rangle_\tau = \sum_{j=1}^N \langle V_j \rangle_\tau,
\end{equation}
where $\langle...\rangle_\tau$ denotes a time average and $V_j$ is
obtained from the Josephson relation,
\begin{equation}
\frac{V_j}{RI_c} = \frac{\hbar}{qRI_c}\frac{d\gamma_j}{dt}
= \frac{1}{Q_J}~\dot{\gamma}_j. \label{eq:joseph}
\end{equation}

A striking feature of this plot is the {\em self-induced resonant
steps} (SIRS), at which $\langle V \rangle_\tau$ remains approximately
constant over a range of applied current.  For this particular choice
of parameters and initial conditions, we see these steps at $\langle V
\rangle_\tau/(NRI_c) = n \tilde{\Omega}/Q_J$.  These steps corresponds
to voltages at which the condition
\begin{equation}
2e\langle V_j \rangle_{t} = n\hbar\Omega
\end{equation}
is satisfied for the individual junctions, with $n = 1$ (upper
horizontal dashed line) and $n = 1/2$ (lower horizontal dashed
line). Thus, the lower step is at 23/80 the voltage of the upper step.
For the latter case, the driving current is smaller than the
retrapping currents of $17$ of the junctions; thus, only $23$ out of
the $40$ junctions are oscillating on this step.  (The retrapping
current is the minimum current for which an underdamped junction is
bistable.)  The steps occur at exactly the voltages where the first
integer and half-integer steps would appear in 
\begin{figure}[t]
\epsfysize=6.9cm
\centerline{\epsffile{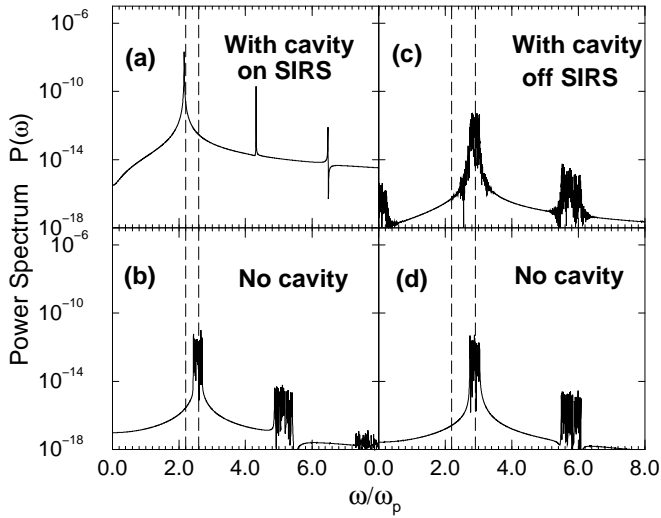}}
\caption{Power spectrum, $P(\omega)$ (Eq.\ \ref{eq:spectrum}), of the
         a.\ c. voltage across the array, plotted versus frequency,
         $\tilde{\Omega}$, at two driving currents: (a) and (b) $I/I_c
         = 0.58$, corresponding to the first integer SIRS, and (c) and
         (d) $I/I_c = 0.65$, slightly off a SIRS.  Other parameters
         are the same as in Fig.\ \ref{fig:IV_cavity}.  Panels (b) and
         (d) are the same as (a) and (c) except that the effective
         coupling to the resonant cavity, $\tg = 0$.  In each panel,
         the left vertical dashed line shows the resonant frequency of
         the cavity, and the right vertical, dashed line shows the
         average resonant frequency of the junctions for the case of
         no coupling to the cavity.}
\label{fig:power}
\end{figure}
\noindent
these junctions, if the junctions were driven by an a.\ c. current of
frequency $\tilde{\Omega}$.  Thus, the radiated energy in the cavity
seems to behave like an a.\ c. drive which acts back to induce these
steps in the junctions of the array.  Similar steps were seen
experimentally in a {\em two-dimensional} array of underdamped
Josephson junctions coupled to a resonant cavity \cite{barbara}, and
in more recent experiments in 1D arrays \cite{vasilic}.

Fig.\ \ref{fig:IV_cavity} also shows the time-averaged scaled total
energy, $\tilde{E}$, contained in the cavity, [right-hand scale of the
Figure].  $\tilde{E}$ is defined as
\begin{equation}
\tilde{E} = \langle\tilde{a}_R^2 + \tilde{a}_I^2\rangle_\tau 
	= g~ \langle a_R^2 + a_I^2 \rangle_\tau = \frac{g}{\hbar \Omega} E,
\label{eq:scaleng}
\end{equation}
where $E = \langle H_{photon} \rangle_\tau$ is the cavity energy; it
is plotted as a function of $I/I_c$ for the same array.  As is
evident, $\tilde{E}$ increases dramatically when the array is on a
SIRS, and is very small otherwise.  This sharp increase signals the
onset of coherence within the array, and can be qualitatively
understood from the equations of motion.  Specifically, when the array
sits on one of the integer SIRS, all the junctions are oscillating in
phase.  Hence, the term driving $\tilde{a}_R$ [the right-hand side of
Eq.\ (\ref{eq:ddgam2})], and thus $\tilde{a}_R$ itself, are both
proportional to the number $N_a$ of active junctions.

Before proceeding further, we briefly review the concept of active
junction number $N_a$, as discussed in Refs.\ [12] and [25].  This
concept has meaning only for underdamped junctions.  Such a junction
is bistable and hysteretic in certain ranges of current - that is, it
can have either zero or a finite time-averaged voltage across it,
depending on the initial conditions.  In the present case, $N_a$
denotes the number of junctions (out of $N$ total) which have a finite
time-averaged voltage drop.  It is possible to tune $N_a$ by suitably
choosing the initial conditions, $\gamma_i$ and $\dot{\gamma}_i$, in
simulations\cite{filatrella,almaas}.

Fig.\ \ref{fig:power} shows the calculated voltage power spectrum of
the a.\ c. component of the total voltage across the array
\begin{equation}
P(\omega) = 2~\lim_{T\rightarrow\infty} \left|\frac{1}{T}\int_0^T V (\tau)
            ~e^{i\omega\tau} d\tau \right|^2, \label{eq:spectrum}
\end{equation}
for two values of the driving current: $I/I_c = 0.58$ [Fig.\
\ref{fig:power}(a) and (b)] and $I/I_c = 0.65$ [Fig.\
\ref{fig:power}(c) and (d)]; all other parameters are the same as in
Fig.\ \ref{fig:IV_cavity}. In (a), all the junctions are on the first
SIRS, while in (c), the array is tuned off this step.  In Fig.\
\ref{fig:power}(b) and \ref{fig:power}(d), we show the same case as in
Fig.\ \ref{fig:power}(a) and (c) respectively, except that the
coupling constant, $\tg$, is artificially set equal to zero.  Note
that in \ref{fig:power}(a), the power spectrum has peaks at the scaled
cavity frequency, $\tilde{\Omega}$, and its harmonics.  This is
evidence that the junctions are all oscillating at frequency
$\tilde{\Omega}$.  In case (b), the junctions are still coupled by the
indirect interaction via the cavity, but the power spectrum shows that
the array is not synchronized in this case; instead, the individual
junctions oscillate approximately at their individual resonant
frequencies and their harmonics and subharmonics.  Hence, the power
spectrum has a spread of frequencies, all of which differ from that of
the cavity.  In cases (b) and (d), the junctions are, of course,
independent of one another, and the power spectrum is that of a
disordered one-dimensional Josephson array with no coupling between
the junctions.

\begin{figure}[tb]
\setbox1=\hbox{\leavevmode \epsfysize 2.8 in \epsfbox{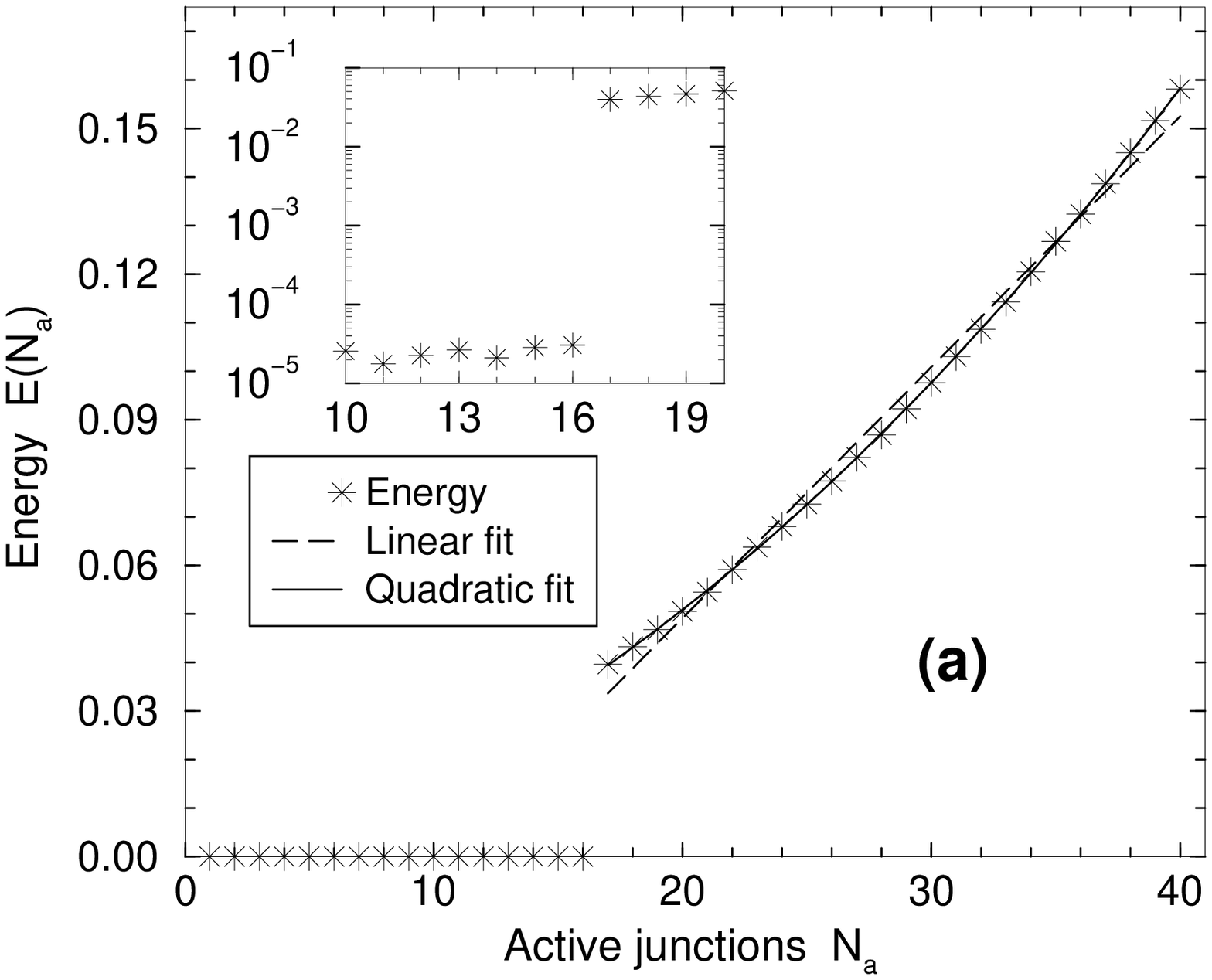}}
\setbox2=\hbox{\leavevmode \epsfysize 2.8 in \epsfbox{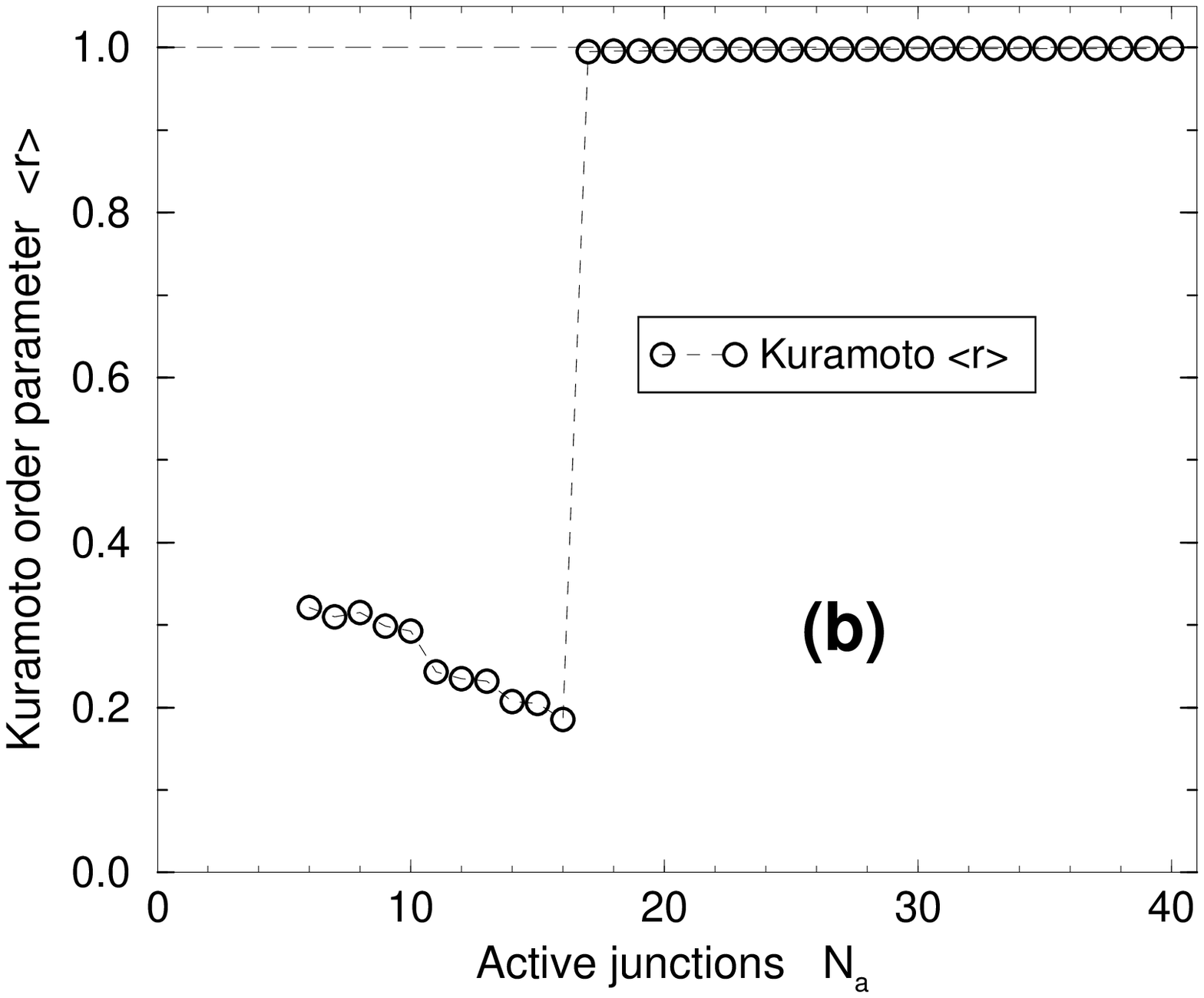}}
\setbox3=\hbox to 17truept{\hfil}
\setbox4=\hbox{\leavevmode \epsfysize 2.4 in \epsfbox{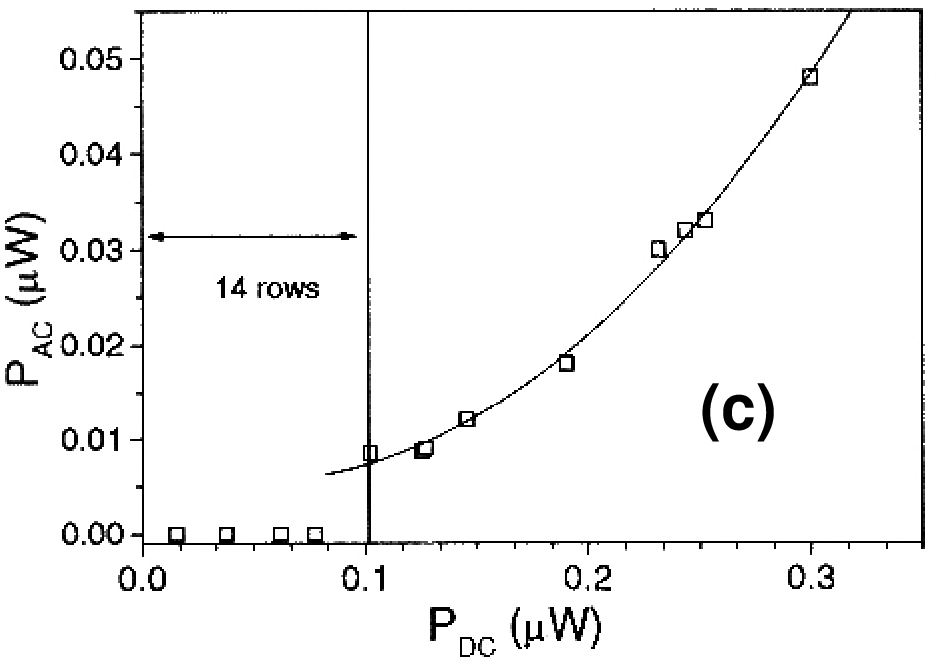}}
\centerline{\hbox{\hsize = 6.5 truein \box4}}
\centerline{\hbox{\hsize = 6.5 truein \box1 \hfil \box3 \hfil \box2}}
\caption{(a) Asterisks: Scaled photon energy $\tilde{E} = g
         E/(\hbar\Omega)$ in the resonant cavity when the array is
         current driven on a SIRS, plotted versus number of active
         junctions, $N_a$.  The array parameters are $N=40$,
         $\tilde{\Omega} = 2.2$, $Q_J = \sqrt{20}$, $\Delta = 0.05$,
         $\tg = 0.001$, and $I/I_c = 0.58$ [cf. Fig.\ \ref{fig:power}
         (a)]. Full curve shows the best fit of $\tilde{E}$ to the
         function $c_2 N_a^2 + c_1 N_a + c_0$ for $N_a > 17$, the
         threshold for synchronization.  The fitting parameters are
         $c_0 = -0.00163$, $c_1 = 0.00125$, and $c_2 = 6.868\cdot
         10^{-5}$. We contrast this fit to the best linear fit (dashed
         line). Inset: $\tilde{E}(N_a)$ near $N_c = 17$, showing jump
         near synchronization threshold.  (b) Open circles: Kuramoto
         order parameter, $\langle r \rangle_\tau$
         [Eq. (\ref{eq:kuramoto})], for the same array.  Dots
         connecting circles are guides to the eye.  The sharp increase
         in $\langle r \rangle_\tau$ and the quadratic increase in
         $\tilde{E}$ both begin at $N_c = 17$. (c) Measured a.\
         c. power as a function of the input d.\ c. power, as obtained
         in Ref.\ [17] for a $3 \times 36$ array. The d.\ c. power is
         proportional to the number of active rows in their array,
         while the a.\ c. power is proportional to the energy
         $\tilde{E}$ in the cavity.  }
\label{fig:energy40}
\end{figure}

We have also calculated the response of a disordered array ($\Delta =
0.05$) of fixed length ($N = 40$ junctions), and a driving current
$I/I_c = 0.58$, when the number of active junctions, $N_a$ is varied.
This current not only lies well within the bistable region, but also
leads to a voltage on the first integer SIRS.  In Fig.\
\ref{fig:energy40}(a), we plot the time-averaged scaled energy of the
cavity, $\tilde{E}(N_a)$ [Eq.\ (\ref{eq:scaleng})], as a function of
$N_a$.  For $N_a < 17$, the active junctions are unsynchronized, and
$\tilde{E}$ is correspondingly small and only weakly dependent on
$N_a$.  There is a sudden jump in $\tilde{E}$ at a critical number of
active junctions $N_c = 17$.  Above this value $\tilde{E}$ increases
as a quadratic function of $N_a$, and we have fitted $\tilde{E}(N_a)$
to the form $\tilde{E} = c_0 + c_1 N_a + c_2 N_a^2$.  The constants
which give the best fit are $c_0 = -0.00163$; $c_1 = 0.00125$; and
$c_2 = 6.868 \cdot 10^{-5}$.  This curve is shown as a full line in
Fig.\ \ref{fig:energy40}(a); the fit is clearly excellent.  As a
contrast, we also show the best linear fit to the same data set
(dashed line); the fit is plainly less good. The magnitude of the jump
in $\tilde{E}$ at $N_a = N_c$ is nearly a factor of $\sim 10^3$, as
shown in the inset to figure.

To measure the degree of synchronization among the Josephson
junctions, we have also calculated the {\em Kuramoto order parameter}
\cite{kuramoto}, $\langle r \rangle_\tau$, for the same parameters,
\newpage
\widetext
\narrowtext \noindent
as a function of number of active junctions, $N_a$.  $\langle r
\rangle_\tau$ is defined by
\begin{equation}
\langle r \rangle_\tau = \langle | \frac{1}{N_a} \sum_{j=1}^{N_a} 
     \exp(i\gamma_j)| \rangle_\tau. \label{eq:kuramoto}
\end{equation}
The results are shown in Fig.\ \ref{fig:energy40}(b).  Note that
$\langle r \rangle_\tau = 1$ represents perfect synchronization among
the active junctions, while $\langle r \rangle_\tau = 0$ would
correspond to no correlations between the different phase differences,
$\phi_i$.  Just as for $\tilde{E}(N_a)$, there is an abrupt increase
in $\langle r \rangle_\tau$ at $N_a = N_c$, indicative of a {\em
dynamical transition} from an unsynchronized to a synchronized state
(with all active junctions locked to the same frequency and having a
common phase), as $N_a$ is increased keeping all other parame-
\vspace*{16.75cm}

\noindent
ters fixed.  As with similar transitions in other models
\cite{strogatz}, this transition is not inhibited by the finite
disorder in the $I_c$'s.  Instead, $\langle r \rangle_\tau$ approaches
unity, representing perfect synchronization. $\langle r\rangle_\tau$
remains finite even for $N_a < N_c$, because even in this regime there
is still some residual correlation among the phases in different
active junctions.  This transition is the dynamic analog of that
analyzed by an equilibrium mean-field theory in Ref.\ [24].

Finally, in Fig.\ \ref{fig:energy40}(c), we show an experimental plot
of the detected a.\ c. power as a function of the input d.\ c.  power,
as measured by Barbara {\it et al}\cite{barbara} for a $3 \times 36$
array.  These quantities are, of course, not equivalent to the
calculated results which are plotted in Fig.\ \ref{fig:energy40} (a).
The input d.\ c. power is equal to the power dissipated in the active
junctions; so it is proportional to $N_a$.  The detec-

\begin{figure}[t]
\epsfysize=6.9cm
\centerline{\epsffile{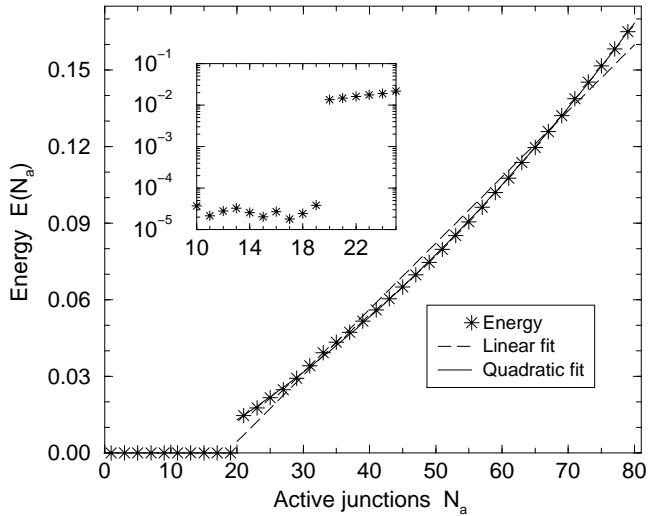}}
\caption{Same as Fig.\ \ref{fig:energy40}(a), except that $N = 80$.
         In this case, the synchronization threshold is $N_c = 20$,
         and the quadratic fit to the energy above synchronization has
         different fitting parameters: $c_0 = -0.01576$, $c_1 =
         0.001149$, and $c_2 = 1.441 \cdot 10^{-5}$.}
\label{fig:energy80}
\end{figure}
\noindent
ted a.\ c. power is that measured by a pickup junction in the cavity,
and thus should be proportional to $\tilde{E}(N_a)$ in our notation.
Despite the differences, our calculated plot (for a one-dimensional
array) appears strikingly similar to their measured plot, especially
as regards the discontinuity at the threshold and the quadratic
dependence on $N_a$ for $N_a$ above the threshold.

In Fig.\ \ref{fig:energy80}, we show the synchronization transition
for an array of $N=80$ junctions, keeping the other parameters the
same as in Fig.\ \ref{fig:energy40}(a).  In this case, the critical
threshold is $N_c = 20$, somewhat larger than for the $N=40$ junction
array. The inset shows that the cavity energy still has a
discontinuity by a factor of $\sim 10^3$. However, the quadratic
function which best fits $\tilde{E}(N_a)$ for $N_a \ge N_c$ is now
described by the different fitting parameters: $c_0 = -0.01576$, $c_1
= 0.001149$, and $c_2 = 1.441 \cdot 10^{-5}$.  Thus, the total length
of the array alters the details but not the qualitative features of
$\tilde{E}(N_a)$.

\begin{figure}[t]
\epsfysize=6.8cm
\centerline{\epsffile{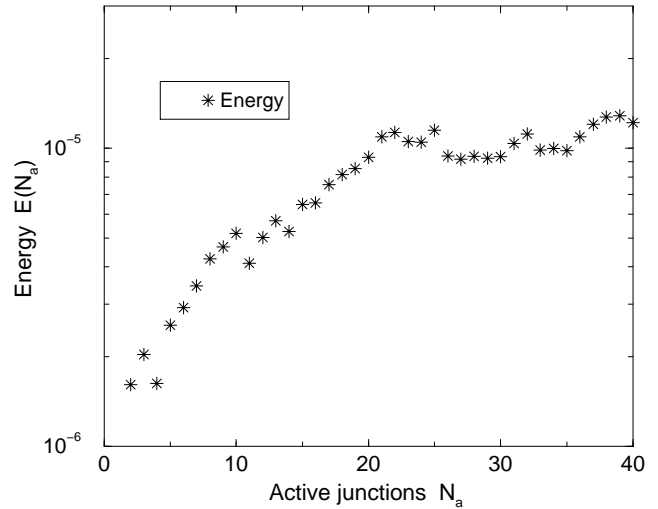}}
\caption{Total scaled cavity energy $\tilde{E}$ as a function of the
         number $N_a$ of active junctions, for the same array
         parameters as in Fig.\ \ref{fig:energy40}(a) except that the
         current is tuned {\em off} any self-induced resonant step:
         $I/I_c = 0.65$ [cf.\ Fig.\ \ref{fig:power}(c)].  In this
         case, $\tilde{E}$ does not increase quadratically with $N_a$
         above a critical threshold; instead, it shows no threshold
         behavior, is only weakly dependent on $N_a$, and is much
         smaller than in Fig.\ \ref{fig:energy40}(a).}
\label{fig:energy_off}
\end{figure}

These calculations were carried out for an array tuned to the first
SIRS.  If, instead, we carry out the same calculation when the array
is tuned to the bistable region but {\em not} tuned to a SIRS, we find
that $\tilde{E}$ does {\em not} increase quadratically with $N_a$.
Instead, $\tilde{E}(N_a)$ shows {\em no} threshold behavior, and,
indeed, varies little with $N_a$. A plot of $E(N_a)$ in this case is
shown in Fig.\ \ref{fig:energy_off}. The parameters are the same as
for the calculation in Fig.\ \ref{fig:energy40}(a), except that the
driving current in this case is $I/I_c = 0.65$, which is not on a SIRS
[cf. Figs.\ \ref{fig:IV_cavity} and \ref{fig:power} ].

\subsection{Effects of Varying the Number of Active Junctions}

\begin{figure}[tb]
\setbox1=\hbox{\leavevmode \epsfysize 2.7 in \epsfbox{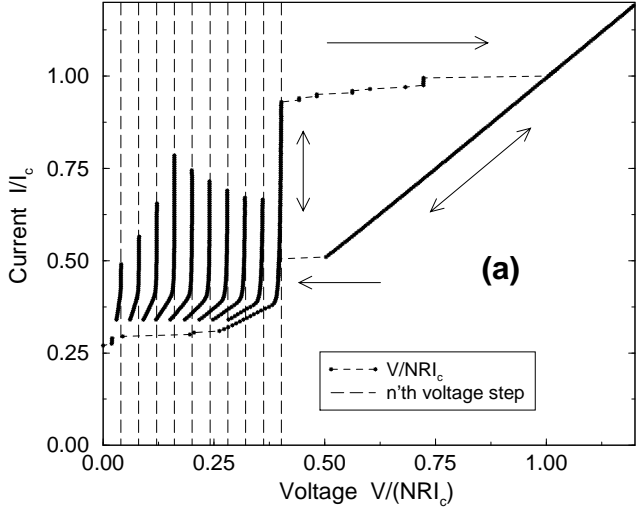}}
\setbox2=\vbox{
	\hbox{\leavevmode \epsfysize 2.6 in \epsfbox{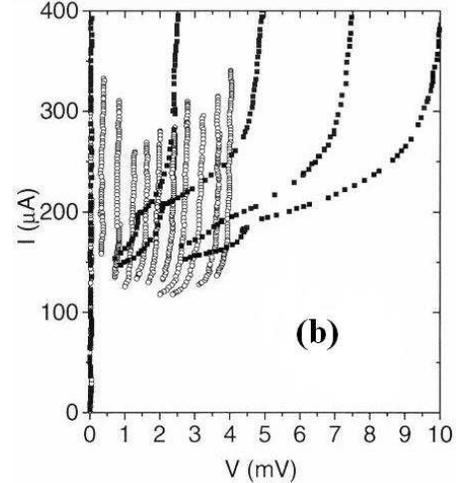}}
 	\hbox to 10truept{\hfil}}
\setbox3=\hbox to 70truept{\hfil}
\setbox4=\hbox to -40truept{\hfil}
\centerline{\hbox{\hsize = 6.5 truein \box4 \box1 \hfil \box3 \hfil \box2 \hfil}}
\caption{(a) Current $I/I_c$ versus time-averaged voltage $\langle V
         \rangle_\tau/(NRI_c)$ for an array containing $N=10$
         junctions, and with damping parameter $Q_J=\sqrt{20}$,
         disorder $\Delta = 0.05$, cavity coupling $\tg = 0.003$, and
         cavity resonance frequency of $\tilde{\Omega} = 1.8$. By
         properly choosing the initial conditions, one can select the
         number $N_a$ of active junctions to be any integer between
         $0$ and $10$.  Each vertical line segment corresponds to a
         portion of the $IV$ characteristic for a particular choice of
         $N_a$, as obtained with increasing current (although the same
         result would be obtained with decreasing current).  The ohmic
         (straight diagonal line) segment is found for $N_a = 10$ with
         {\em decreasing} current.  The dashed vertical lines indicate
         the voltages of the expected integer SIRS's.  The dashed,
         nearly horizontal line corresponds to increasing the voltage
         on the $N_a = 10$ IV characteristic; the dots and very short
         vertical line segments within this dashed line corresponds to
         currents at which several of the active junctions jump to the
         $n = 2$ SIRS.  The short, nearly horizontal dashed line in
         the lower left-hand corner occurs on the decreasing current
         branch with $10$ active junctions.  The very short vertical
         line segments within this dashed region correspond to several
         active junctions synchronizing on the $n = 1/2$ SIRS, while
         the remainder are in the state of $\langle V_j \rangle_\tau =
         0$.  (b). Measured IV characteristics for a $3 \times 36$
         array \cite{barbara}.  The open circles represent
         self-induced resonant steps corresponding to different
         numbers of active rows.  Full squares are believed to be
         examples of resistance steps\cite{wenbin}.}
\label{fig:SIRS_10}
\end{figure}

In Fig.\ \ref{fig:SIRS_10}(a), we show a series of IV characteristics
for a 10-junction array ($N = 10$), calculated by varying the 

\noindent
number $N_a$ of active junctions from $1$ to $10$.  Each solid
vertical line segment corresponds to the IV characteristic for a {\em
different} $N_a$, and represents $N_a$ junctions sitting on the first
integer SIRS.  The width of each segment represent the current height
for that step, as found in our calculation.  The dashed vertical lines
show the expected voltages for the integer SIRS's, and are good
matches for the calculated voltages for the various $N_a$'s.  The long
straight diagonal line segment, which is common to all the different
$N_a$'s, represents the ohmic part of the IV characteristic with all
junctions active. The nearly horizontal dashed line in the upper right
hand corner of the Figure shows the IV characteristic for increasing
voltage with $N_a = 10$.  The very short vertical segments within this
dashed line correspond to several junctions which have been excited to
{\em higher} steps, specifically the $n = 2$ (second integer step)
while the remaining junctions are on the $n = 1$ step.  The horizontal
dashed line on the lower left represents the low-voltage end of the
$N_a = 10$ IV characteristic (on decreasing current).  The short
vertical segment within this dashed line corresponds to {\em
fractional} SIRS's -- specifically, three of the junctions have
slipped from the $n = 1$ to the $n = 1/2$ step, while the rest are in
the $\langle V_j \rangle_\tau = 0$ state (the driving current is
smaller than their individual retrapping currents).  Thus, we see both
the higher integer and the fractional SIRS's in these one-dimensional
arrays.

In Fig.\ \ref{fig:SIRS_10}(a), although we show the full hysteresis
loop only for $N_a = 10$, the IV curves for other values of $N_a$ are
also hysteretic.  In all cases for which $N_c \leq N_a < 10$, the
number of active junctions increases when the SIRS becomes unstable,
and individual junctions jump into the 

\newpage
\widetext
\narrowtext
\noindent
$n = 2$ SIRS state; ohmic behavior is not attained until $I/I_c >1$.
For $N_a < N_c$, the array behaves somewhat differently: when the SIRS
becomes unstable, $N_a$ is unchanged, and the IV curve immediately
becomes ohmic.  When $I/I_c \sim 1$ in this regime, the remaining
junctions become active and the IV characteristic also becomes ohmic.
For this particular array, $N_c = 4$.

As a comparison, we also show, in Fig.\ \ref{fig:SIRS_10}(b), the IV
characteristics as measured for a $3 \times 36$ underdamped array, by
Barbara {\it et al}\cite{barbara}.  The open circles correspond to the
steps observed for different numbers of active rows (from $1$ to $10$
in this instance), which are produced when an in-plane magnetic field
reduces the critical current of the individual junctions.  The more
widely spaced dark rows are believed to be examples of resistance
steps \cite{wenbin}.  The steps (open circles) very much resemble
those of Fig.\ \ref{fig:SIRS_10}(a), even including the low-current
falloff (though the shapes of the curves are slightly different).

\begin{figure}[tb]
\setbox1=\hbox{\leavevmode \epsfysize 2.7 in \epsfbox{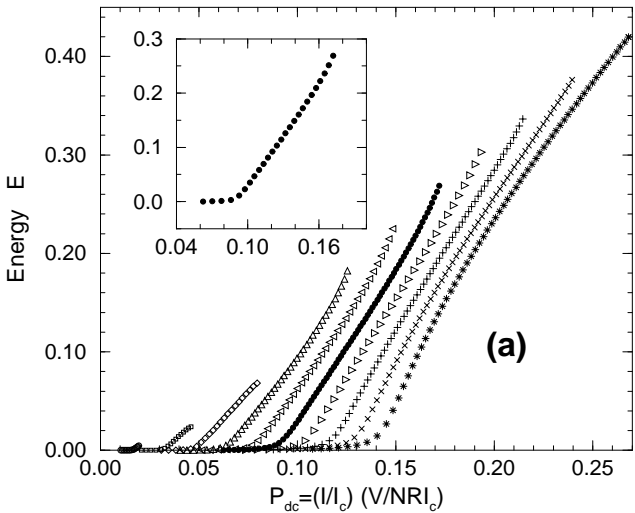}}
\setbox2=\vbox{\hbox{\leavevmode \epsfysize 2.6 in \epsfbox{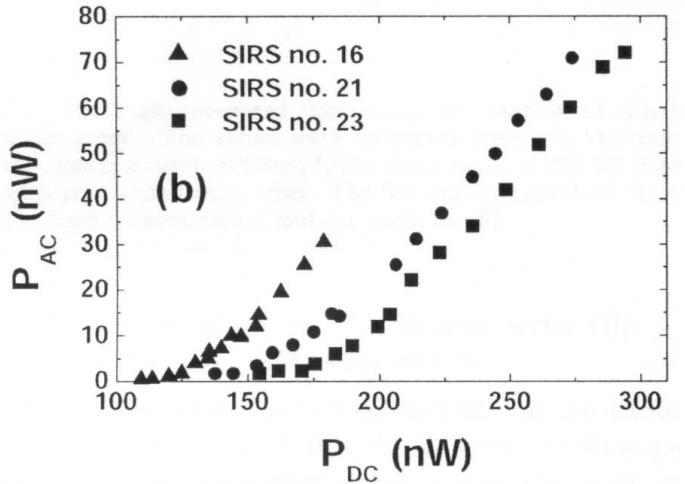}} \hbox to 20truept{\hfil}}
\setbox3=\hbox to 25truept{\hfil}
\centerline{\hbox{\hsize = 6.5 truein \box1 \hfil \box3 \hfil \box2}}
\caption{(a) Calculated total energy $\tilde{E}$ within the cavity,
         plotted versus d.\ c. power, $P_{dc}$, for $N_a$ active
         junctions synchronized on the $n = 1$ SIRS, for an array of
         10 junctions ($N = 10$), using the same parameters as in
         Fig.\ \ref{fig:SIRS_10}.  Each curve segment corresponds to a
         different value of $N_a$ between $1$ (leftmost curve) and
         $10$ (rightmost curve).  $P_{dc} = (I V)/(N R I_c^2)$
         represents the power per junction fed into the array by the
         d.\ c.  current. For each $N_a$, the curve segment ends when
         the array leaves the SIRS. Note that the active junctions in
         the array synchronize on the SIRS when $N_a \ge 4$,
         i.~e. $N_c = 4$ for this array.  Inset: An enlargement of the
         calculated curve for $N_a = 6$ (filled circles). (b)
         Experimental results for a $4 \times 36$ array as reported in
         Ref.\ [43].  From left to right, these results correspond to
         $N_a=16$, $N_a=21$, and $N_a=23$ active rows (all in the
         coherently radiating state with $N_a > N_c$).}
\label{fig:steppower}
\end{figure}

We have also calculated $\tilde{E}$, the energy in the cavity, as a
function of injected d.\ c. power, $P_{dc}$, when the array is biased
on a SIRS, for several choices of array parameters.  A typical example
of our results is shown in Fig.\ \ref{fig:steppower}(a), where
$\tilde{E}$ is plotted versus $P_{dc} \equiv (I/I_c) [\langle V
\rangle_\tau / (N R I_c)]$ for an array of ten junctions, using the
same parameters as in Fig.\ \ref{fig:SIRS_10}(a) and varying the
values of $N_a$.  Each curve corresponds to a different number $N_a$
of active junctions, and, for each $N_a$, we sweep current across the
$n = 1$ SIRS 
\vspace*{11.9cm}

\noindent
(leftmost curve corresponds to $N_a = 1$, and rightmost to $N_a =
10$).  The curves end when the SIRS's become unstable.  Each curve is
quadratic at low $P_{dc}$ and approximately linear at higher $P_{dc}$.
For comparison we also show the corresponding {\em experimental} plots
\cite{asc2000} for a $4 \times 36$ array for $N_a = 16$, $21$, and
$23$ active rows [Fig.\ \ref{fig:steppower}(b)].  In all cases the
experimental array is above $N_c$, the coherence threshold.  The
similarity between the experimental and calculated curves is
strikingly apparent.

\subsection{Effects of Changing Model Parameters}

Finally, we have studied how our numerical results depend on the
parameters of our model.  There are several parameters of interest:
the number of junctions $N$, the disorder parameter $\Delta$, the
damping parameter $Q_J$, the coupling constant $\tilde{g}$, and the
normalized cavity mode frequency $\tilde{\Omega}$. Clearly, a thorough
numerical investigation of all these parameters is out of the
question.  We have therefore varied only two parameters in the present
paper: $Q_J$ and $\tilde{g}$.

\begin{figure}[tb]
\setbox1=\hbox{\leavevmode \epsfysize 2.9 in \epsfbox{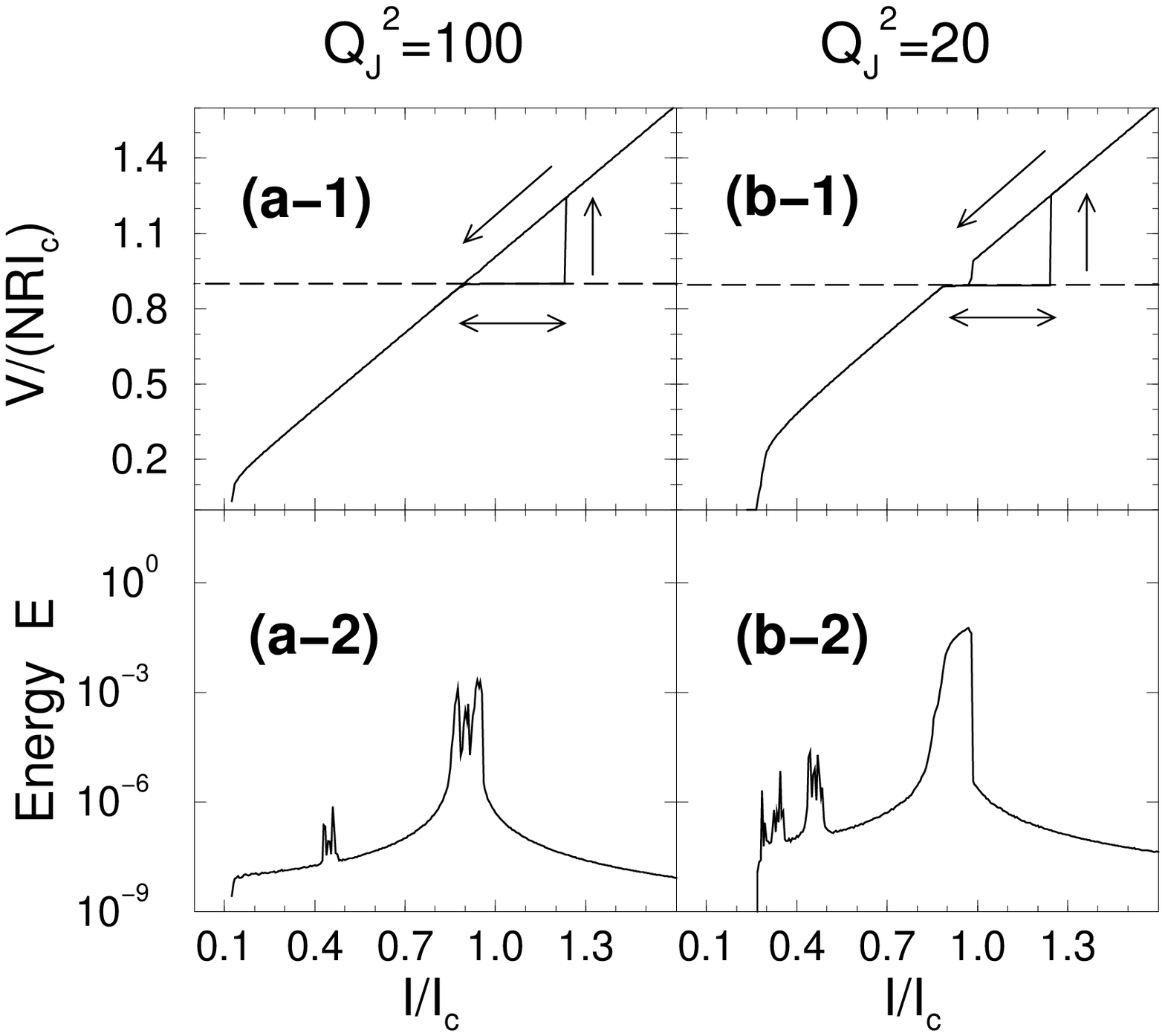}}
\setbox2=\hbox{\leavevmode \epsfysize 2.9 in \epsfbox{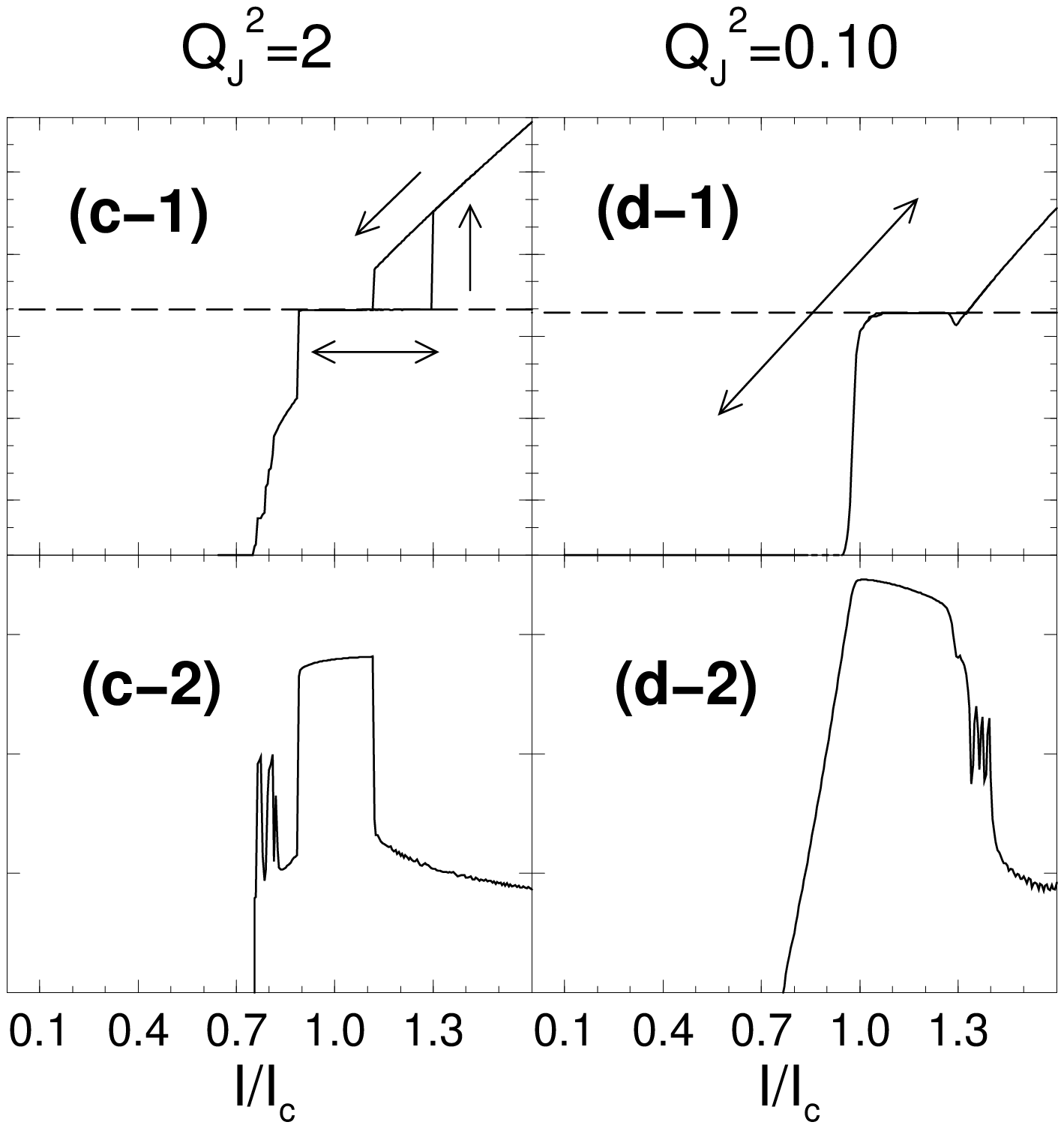}}
\setbox3=\hbox{\leavevmode \epsfysize 2.9 in \epsfbox{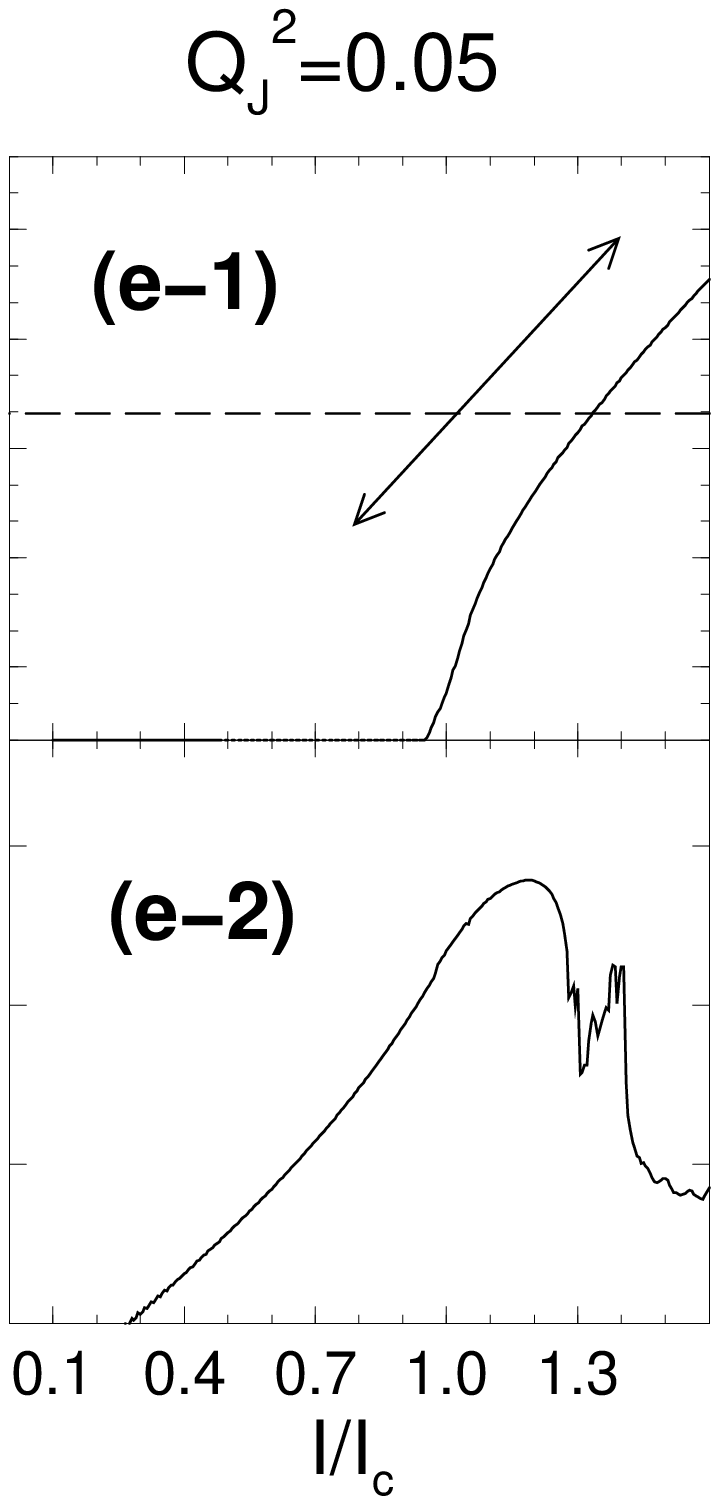}}
\centerline{\hbox{\hsize = 6.5 truein \box1 \box2 \box3}}
\caption{This Figure illustrates the effects of changing the damping
         parameter $Q_J$ while holding other array parameters fixed.
         Note that (d) and (e) correspond to {\em overdamped}
         junctions.  Panels (a) - (e) show results for $Q_J^2 = 100$,
         $20$, $2$, $0.1$, and $0.05$ for an array of 10 junctions ($N
         = 10$), with coupling strength $\tilde{g} = 4. \times
         10^{-4}$ and disorder parameter $\Delta = 0.05$.  In all
         cases, the cavity resonant frequency is chosen such that the
         expected voltage for the SIRS is $\tilde{\Omega}/Q_J = 0.9$.
         The top panels show the time averaged voltage $\langle V
         \rangle_\tau/(NRI_c)$ across the array as a function of
         driving current $I/I_c$. Note the absence of clear hysteresis
         in (d) and (e), which correspond to overdamped junctions. The
         arrows indicate whether the trace is calculated for
         increasing or decreasing current.  Lower panels show the
         time-averaged total energy $\tilde{E} = \langle a_R^2 + a_I^2
         \rangle_\tau$ in the cavity, calculated as a function of {\em
         decreasing} current only.}
\label{fig:changeQ}
\end{figure}

Fig.\ \ref{fig:changeQ} shows the total time-average voltage $\langle
V \rangle_\tau$ across the array, and the total time-averaged energy
$\tilde{E}$ in the array as a function of driving current $I/I_c$, for
$Q_J = \sqrt{100}$, $\sqrt{20}$, $\sqrt{2}$, $\sqrt{0.1}$, and
$\sqrt{0.05}$, all for $\tilde{g} = 4 \times 10^{-4}$, $N = 10$, and
$\Delta = 0.05$.  In each case, the resonant frequency of
\newpage
\widetext
\narrowtext
\noindent
of the cavity $\tilde{\Omega}$ is chosen such that the scaled voltage
$\tilde{\Omega}/Q_J = 0.9$.  This choice insures that the voltage lies
within the bistable region of the $IV$ characteristic for the
underdamped junctions.  The arrows in the upper panel indicate the
direction in which the current is swept. We show only the energy in
the cavity for the decreasing current branch.

Several features of these curves are apparent.  First, the SIRS's are
wider on the increasing than the decreasing branches.  For the most
underdamped case (a), there are no visible SIRS's on decreasing the
current.  Secondly, the cavity energy shows clear signs of a resonant
interaction between the array and the cavity in cases
(a)-(c). Finally, there are strong indications of an integer SIRS even
for the overdamped case (d), where there is no bistable region in the
uncoupled IV characteristics.  [We find an even clearer integer SIRS
in case (d) if we increase $\tilde{g}$ by a factor of 10. In this
case, a SIRS also develops in (e) (not shown in the Figure)].

\begin{figure}[tb]
\setbox1=\hbox{\leavevmode \epsfysize 3.1 in \epsfbox{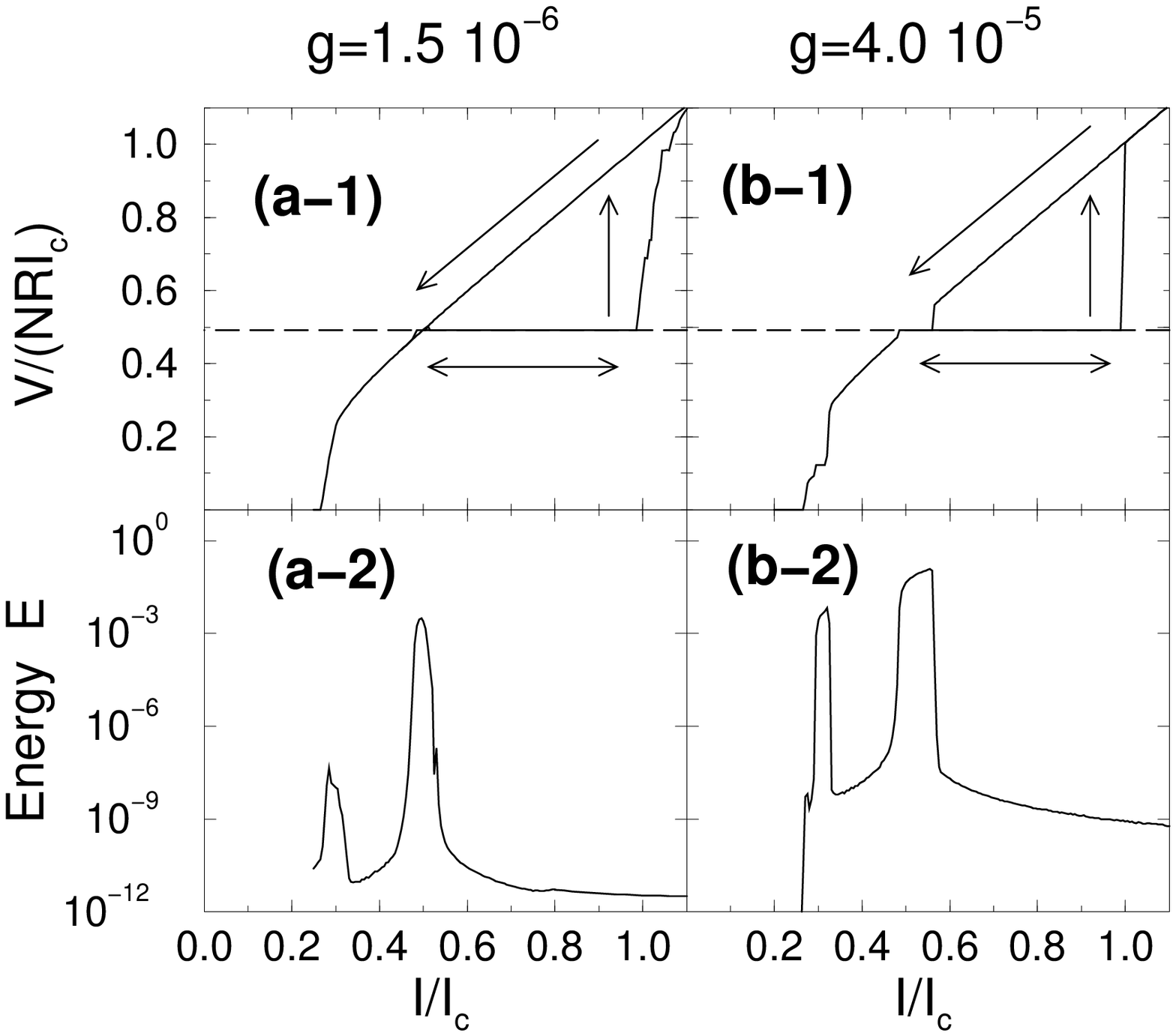}}
\setbox2=\hbox{\leavevmode \epsfysize 3.1 in \epsfbox{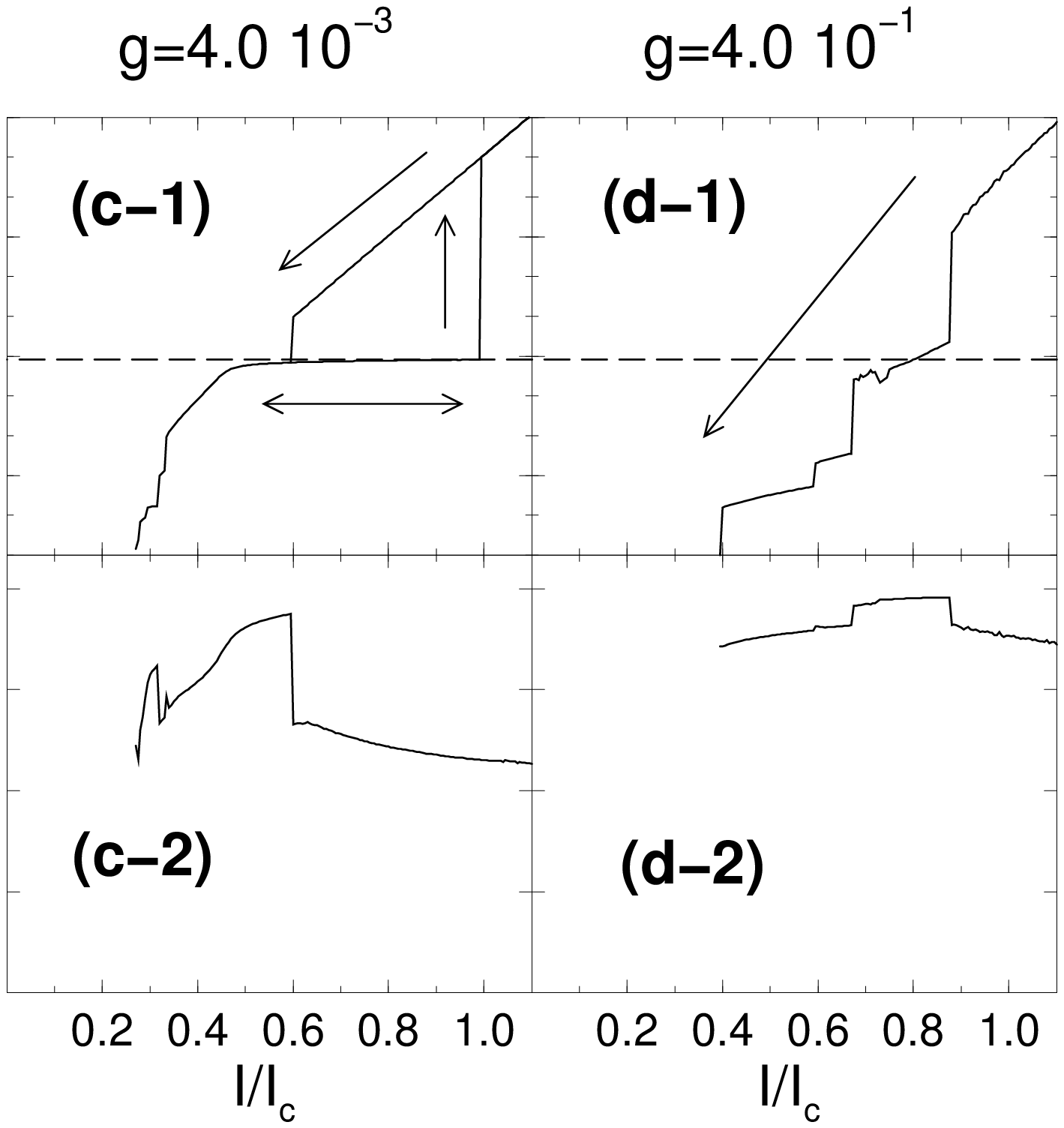}}
\centerline{\hbox{\hsize = 6.5 truein \box1 \box2}}
\caption{This Figure illustrates the effects of changing the coupling
         parameter $\tilde{g}$ while holding the other parameters
         fixed.  Panels (a) - (d) correspond to $\tilde{g} = 1.5\times
         10^{-6}$, $4 \times 10^{-5}$, $4 \times 10^{-3}$, and $4
         \times 10^{-1}$, all with $Q_J = \sqrt{20}$, $N = 10$, and
         $\Delta = 0.05$.  Top panels: Time-averaged total voltage
         across the array, $\langle V \rangle_\tau/(NRI_c)$, versus
         current $I/I_c$.  Arrows indicate the direction of current
         sweep.  Bottom panels: total time-averaged energy $\tilde{E}$
         in the cavity as a function of $I/I_c$, all calculated for
         {\em decreasing} current bias.}
\label{fig:changeg}
\end{figure}

In Fig.\ \ref{fig:changeg} (a) - (d), we plot $\langle V \rangle_\tau$
and cavity energy $\tilde{E}$ versus $I/I_c$ for several values of the
coupling constant $\tilde{g}$, all for $Q_J = \sqrt{20}$, $N = 10$,
$\Delta = 0.05$, and $\tilde{\Omega}/Q_J = 0.9$.  Once again, the
arrows in the upper panels denote direction of current sweep.  As
discussed in the next section, we believe that experiments have been
carried out for $\tilde{g}$ somewhere in the range of panels (a) and
(b).  For (a), there is a very wide first integer SIRS on the upward
sweep but none visible the downward direction.  In (b) and (c), there
are SIRS's in both directions, but wider on the upward sweep.  In case
(d), which we show for completeness but believe to correspond to an
\newpage
\vspace*{10.05cm}

\noindent
unattainable large coupling, there are no detectable steps but several
discontinuities in the IV characteristic which are discussed below.
The cavity energy $\tilde{E}$ is calculated on the {\em decreasing}
sweep.  It shows a resonant enhancement even when the IV's (on this
downward sweep) show no indication of a SIRS.  [This enhancement is
also visible on the upward sweep, which we have not shown.]  In panel
(a), $\tilde{E}$ shows a resonance at a current corresponding to a
half-integer SIRS, but the IV characteristics themselves show no clear
evidence of such a SIRS.  In cases (b) and (c), we find that at these
currents some fraction of the junctions have phase-locked onto the $n
= 1/2$ step while the others are in the $\langle V_j\rangle_\tau = 0$
state.  Another noteworthy feature is that as $\tilde{g}$ increases,
the integer steps in Fig.\ \ref{fig:changeg} (a) - (d) acquire a
noticeable nonzero slope, and also become more and more rounded near
their lower edge.

In order to shed some light on the IV characteristics of Fig.\
\ref{fig:changeg} (d), we have looked at the $\langle V_i
\rangle_\tau$'s across the individual junctions. Depending on $I/I_c$,
all the $\langle V_i \rangle_\tau$'s may be different, they may all be
equal, or they fall into two or three groups.  For certain $I$'s, some
of the $\langle V_i \rangle_\tau$'s are nonzero while others vanish.
This last behavior presumably arises from the disorder in the critical
currents.

\section{Discussion}

\subsection{Comparison Between Calculated Results and Experiment}

We now compare the present results to experi-ment
\cite{barbara,vasilic,asc2000}.  Most of the published experiments
thus far 
\newpage
\widetext
\narrowtext
\noindent
have been carried out on two-dimensional arrays.  Their main
features include the following:

(a). When the array is driven by a current, the IV characteristics
show self-induced resonant steps.

(b). These steps are reported for any number of active junctions
$N_a$.

(c). Above a critical threshold number $N_c$ of active junctions, the
a.\ c. power output (i. e., the energy in the cavity) increases
quadratically with $N_a$.  When the $N_a$ is increased through the
threshold, the detected a.\ c. power in the cavity jumps by several
orders of magnitude at the threshold.

(d). The array can be experimentally tuned so that different numbers
of rows (i. e., different numbers of active junctions) are on the $n =
1$ SIRS.

(e). When $N_a$ junctions are on a SIRS and the current drive is
varied, the $P_{ac}$ versus $P_{dc}$ curve is quadratic for low
$P_{ac}$ and linear for high $P_{dc}$.

Our numerical results show all five of these features for a {\em one
dimensional} array.  Thus, they suggest that the behavior seen in the
2D experiments should be visible even for a 1D system.  Indeed, a
recent report \cite{vasilic} suggests that all the features (a) - (e)
are indeed experimentally observable in 1D.

We now elaborate on some of these points.  The SIRS's emerge naturally
from our equations of motion [Eqs.\ (\ref{eq:ddgam1}) and
(\ref{eq:ddgam2})]. Another notable point is that we can numerically
control the number of active junctions $N_a$ by tuning the initial
conditions.  This tuning is possible because the junctions are
underdamped and have an applied current regime within which they are
bistable.  The chosen $N_a$ 
\newpage
\vspace*{10.45cm}

\noindent
determines whether the array is above or below the coherence threshold
$N_c$.  If $N_a > N_c$, then we usually find that, when the junctions
lock onto a SIRS, they all lock onto the same, n = 1 step (first
integer step).  The voltage drop across the array is then $\langle V
\rangle_\tau/(NRI_c) = N_a\tilde{\Omega}/Q_J$.  Thus, the same array
can produce an IV characteristic with multiple branches, each
corresponding to a different number of SIRS's.  This behavior is in
agreement with the behavior seen in Ref.\ [17].

If $N_a < N_c$, then our calculations still produce integer SIRS's,
but these steps are not coherent with one another.  That is, although
each junction is individually locked onto the same fundamental
frequency, which is close to the frequency $\tilde{\Omega}$ of the
cavity, the active junctions are out of phase with one another, and
hence do not generate an energy in the cavity which varies
quadratically with $N_a$.  Also, even above the coherence threshold
($N_a > N_c$), if the junctions are not locked on the steps, the array
is not coherent at the coupling constant which produces the steps --
that is, the power spectrum is reminiscent of that of an array of
independent junctions, and does not show a series of multiples of a
single fundamental frequency.  Under these off-step conditions, the
array can be made coherent, but only if the coupling constant is
increased by several orders of magnitude above that needed to produce
the SIRS's.

Under some conditions, our calculations yield not only the first
integer SIRS's but also overtone steps (higher integer steps), and
fractional steps.  The widths of our fractional steps are extremely
small, and the steps are obtainable only by a delicate tuning of the
current, initial 
\newpage
\widetext
\narrowtext
\noindent
conditions, and current sweep rate.  This sensitivity may explain why
these fractional steps have not, as yet, been detected experimentally,
though the overtone steps have been found \cite{private}.

Not only the general features but even some of the details of our
calculations seem to agree well with experiment.  For example, the
results in Fig.\ \ref{fig:steppower}(a) show the variation of a.\
c. power (that is, the electromagnetic energy in the cavity) with the
input d.\ c. power.  The different curves correspond to distinct
number of active junctions, $N_a$, for this particular array. All the
curves show a gradual, nearly parabolic onset but become nearly linear
at higher input power (that is, near the high-current edge of the
step). The main difference between the cases $N_a > N_c$ and $N_a <
N_c$, is the behavior of the energy in the cavity after the SIRS
becomes unstable (for increasing $I/I_c$).  When $N_a < N_c$, we find
that $\tilde{E} \sim 10^{-5}$ at such input powers, while in the
opposite case $\tilde{E} \sim 0.1$.  (This behavior is not shown in
the Figure.)  Very similar behavior to that shown in Fig.\
\ref{fig:steppower}(a) has recently been reported experimentally in
Ref.\ [43], and is shown in Fig.\ \ref{fig:steppower}(b).  The
similarity between the results of Ref.\ [43] and the present work is
apparent.  A related experiment has also been reported in which a 30\%
d.\ c. to a.\ c. conversion rate was achieved \cite{lobbapl}.

\subsection{Qualitative Discussion of Underlying Physics}

We now briefly discuss the physics behind the present numerical
results.  First, the existence of a transition 

\newpage
\vspace*{9.8cm}

\noindent
from incoherence to coherence, as a function of $N_a$, results from
the ``mean-field-like'' nature of the interaction between the
junctions and the cavity.  Specifically, because each junction is
effectively coupled to every other junction via the cavity, the
strength of the effective coupling increases with $N_a$.  Thus, for
any $\tilde{g}$, a transition to coherence is to be expected for
sufficiently large $N_a$.  A similar argument was made in the
equilibrium case in Ref.\ [24].

Above the coherence transition, the self-induced resonant steps can
also be qualitatively understood by referring to the underlying
equations (\ref{eq:ddgam1}) and (\ref{eq:ddgam2}).  When a current is
applied, it sets all the $\gamma_i$'s into motion, according to Eq.\
(\ref{eq:ddgam1}).  If these $\gamma_i$'s all oscillate at the same
fundamental frequency, they act as a driving term which causes
$\tilde{a}_R$, and hence $\ddot{\tilde{a}}_R$, to oscillate at the
same frequency, according to Eq.\ (\ref{eq:ddgam2}).  This
$\ddot{\tilde{a}}_R$ then behaves like an a.\ c.  current drive in
Eq.\ (\ref{eq:ddgam1}).  The combined d.\ c. and a.\ c. drives in Eq.\
(\ref{eq:ddgam1}) produce SIRS's, just as a combined d.\ c. and a.\
c. current produce Shapiro steps in a conventional Josephson junction.
This same picture also makes it clear why the cavity energy increases
quadratically with $N_a$ above the threshold: in this regime, the
``inhomogeneous'' term on the right-hand side of Eq.\
(\ref{eq:ddgam2}) is proportional to $N_a$ and, therefore, so is
$a_R$.  The whole process occurs self-consistently because the two
equations are coupled.  The effective ``a.\ c. driving current''
$\ddot{\tilde{a}}_R$ in Eq.\ (\ref{eq:ddgam1}) is also proportional to
$N_a$.  Since the height of the first integer Shapiro step in a
conventional junction is proportional to $J_1 (\alpha I_{ac})$ where
$I_{ac}$ is the amplitude of the a.\ c.  driving current and $\alpha$
is a constant related to the frequency, one might expect that the
width of the SIRS's would have an oscillatory dependence on $N_a$.
There are some slight hints of this behavior in our numerical results
[cf.\ Fig.\ \ref{fig:SIRS_10}(a)].

This description also suggests why the steps occur even in
one-dimensional arrays.  Their occurrence depends, not on the
dimensionality of the array, but only on the existence of a suitable
induced a.\ c. drive.  Indeed, such steps have recently been reported
in 1D arrays \cite{vasilic}, consistent with the present model.  The
in-plane magnetic field used in the earlier experiments is apparently
needed only to lower the Josephson critical currents sufficiently that
the resonant frequency $\Omega$ occurs in the bistable region of the
IV characteristics.

All the numerical results in the present paper are obtained in the
``semi-classical'' regime, where the various operators are regarded as
$c$-numbers.  It would be of interest to study the array dynamics of
the array in the quantum regime, where the number of photons is small.
A recent numerical study of this kind (but only for the equilibrium
properties) has been carried out for a SQUID in a resonant cavity
(without resistively-shunted damping) \cite{squid}.

In summary, we have derived the Heisenberg equations of motion for a
model Hamiltonian which describes a one-dimensional array of
underdamped Josephson junctions coupled to a resonant cavity.  We have
numerically solved these equations in the classical limit, valid in
the limit of large numbers of photons in the cavity.  In the presence
of a d.\ c. current drive, we find numerically that (i) the array
exhibits self-induced resonant steps (SIRS), similar to Shapiro steps
in conventional arrays; (ii) there is a transition between an
unsynchronized and a synchronized state as the number of active
junctions is increased while other parameters are held fixed; and
(iii) when the array is biased on the first integer SIRS, the total
energy increases quadratically with number of active junctions.  Our
results are in quite detailed agreement with experiment, even
though the experiments are largely carried out in 2D.  Thus, the
present model strongly suggests that a 2D array is not necessary in
order to obtain the observed SIRS's.  The results also strongly suggest 
that the experimental data considered here can be understood in 
terms of a model involving {\em strictly classical} equations of motion,
without the necessity of introducing new, non-classical physics.

\section{Acknowledgments.}

We are most grateful for support from NSF through grants DMR97-31511
and DMR01-04987.  Computational support was provided by the Ohio
Supercomputer Center, and the Norwegian University of Science and
Technology (NTNU).  We thank C. J. Lobb, P. Barbara, and B. Vasilic
for useful conversations.


\begin{references}
\bibitem[*]{email1} Electronic address: Almaas.1@osu.edu

\bibitem[\dag]{email2} Electronic address: stroud@mps.ohio-state.edu

\bibitem{booi} See, e. g.\, P. A. A. Booi and S. P. Benz, Appl.
Phys. Lett. {68}, 3799 (1996).

\bibitem{han} S. Han, B. Bi, W. Zhang, and J. E. Lukens, Appl.
Phys. Lett. {\bf 64}, 1424 (1994).

\bibitem{kaplunenko} V. K. Kaplunenko, J. Mygind, N. F. Pedersen,
and A. V. Ustinov, J. Appl. Phys. {\bf 73}, 2019 (1993).

\bibitem{wan} K. Wan, A. K. Jain, and J. E. Lukens, Appl. Phys.
Lett. {\bf 54}, 1805 (1989).

\bibitem{hadley} P. Hadley, M. R. Beasley, and K. Wiesenfeld,
Phys. Rev. B {\bf 38}, 8712 (1988).

\bibitem{octavio} M. Octavio, C. B. Whan, and C. J. Lobb, Appl.
Phys.\ Lett. {\bf 60}, 766 (1992).

\bibitem{wiesenfeld1} K. Wiesenfeld, S. P. Benz, and P. A. A.
Booi, J. Appl. Phys. {\bf 76}, 3835 (1994).

\bibitem{braiman} Y. Braiman, W. L. Ditto, K. Wiesenfeld, and M.
L. Spano, Phys. Lett. A {\bf 206}, 54 (1995).

\bibitem{darula} M. Darula, S. Beuven, M. Siegel, A. Darulova, P.
Seidel, Appl. Phys. Lett. {\bf 67}, 1618 (1995).

\bibitem{whan} C. B. Whan, A. B. Cawthorne, and C. J. Lobb,
Phys. Rev. B {\bf 53}, 12340 (1996).

\bibitem{filatrella1} G. Filatrella, N. F. Pedersen, and K.
Wiesenfeld, Appl. Phys. Lett. {\bf 72}, 1107 (1998).

\bibitem{filatrella} G. Filatrella, N. F. Pedersen, and K.
Wiesenfeld, Phys. Rev. E {\bf 61}, 2513 (2000).

\bibitem{jain} A. K. Jain, K. K. Likharev, J. E. Lukens, and J.
E. Sauvageau, Phys. Rep. {\bf 109}, 309 (1984).

\bibitem{benz} S. P. Benz and C. J. Burroughs, Appl. Phys.
Lett. {\bf 58}, 2162 (1991).

\bibitem{cawthorne1} A. B. Cawthorne, P. Barbara, and C. J. Lobb,
IEEE Trans. Appl. Supercond. {\bf 7}, 3403 (1997).

\bibitem{neutral} K. Wiesenfeld and P. Hadley, Phys. Rev. Lett. {\bf
62}, 1335 (1989); S. Nichols and K. Wiesenfeld, Phys. Rev. A {\bf 45},
8430 (1992); K. Wiesenfeld, S. P. Benz, and P. A. A. Booi,
J. Appl. Phys. {\bf 76}, 3835 (1994).

\bibitem{barbara} P. Barbara, A. B. Cawthorne, S. V. Shitov, and
C. J. Lobb, Phys. Rev. Lett. {\bf 82}, 1963 (1999).

\bibitem{bonifacio} L. A. Lugiato and M. Milani, Nuovo Cimento B {\bf
55} 417 (1980); R. Bonifacio, F. Casagrande, and L. A.  Lugiato,
Opt. Comm. {\bf 36}, 159 (1981); R. Bonifacio, F. Casagrande, and
G. Casati, Opt. Comm. {\bf 40}, 219 (1982); R. Bonifacio,
F. Casagrande, and M. Milani, Lett. Nuovo Cimento {\bf 34}, 520
(1982).

\bibitem{wiesenfeld} K. Wiesenfeld, P. Colet, and S. H. Strogatz,
Phys. Rev. Lett. {\bf 76}, 404 (1996).

\bibitem{cawthorne} A. B. Cawthorne, P. Barbara, S. V. Shitov,
C. J. Lobb, K. Wiesenfeld, and A. Zangwill, Phys. Rev. B {\bf
60}, 7575 (1999).

\bibitem{abdullaev} F. K. Abdullaev, A. A. Abdumalikov, Jr.,
O. Buisson, and E. N. Tsoy, Phys. Rev. B {\bf 62}, 6766 (2000).

\bibitem{fistul} P. Caputo, M. V. Fistul, B. A. Malomed, S. Flach, and
A. V. Ustinov, Phys. Rev. B {\bf 59}, 14050 (1999).

\bibitem{jensen} N.\ Gr{\o}nbech-Jensen, R.\ D.\ Parmentier, and N.\
F.\ Pedersen, Phys. Lett. A {\bf 142}, 427 (1989); N.\
Gr{\o}nbech-Jensen, N.\ F.\ Pedersen, A.\ Davidson, and R.\ D.\
Parmentier, Phys. Rev. B {\bf 42}, 6035 (1990); R.\ Monaco, N.\
Gr{\o}nbech-Jensen, and R.\ D.\ Parmentier, Phys. Lett. A {\bf 151},
195 (1990); G.\ Filatrella, G.\ Rotoli, N.\ Gr{\o}nbech-Jensen, R.\
D.\ Parmentier, and N.\ F.\ Pedersen, J. Appl. Phys. {\bf 72}, 3179
(1992).

\bibitem{harb} J. K. Harbaugh and D. Stroud, Phys. Rev. B {\bf
61}, 14765 (2000).

\bibitem{almaas} E. Almaas and D. Stroud, Phys. Rev. B {\bf 63},
144522 (2001); Phys.\ Rev.\ B{\bf 64}, 179902(E) (2001).  As the
Erratum notes, this paper includes a mistake in the damping and
current-drive terms of the Hamiltonian, which leads to numerical
results inferior to those in the present paper.

\bibitem{slater} J. C. Slater, {\it Microwave Electronics} (D. Van
Nostrand, New York, 1950).

\bibitem{yariv} A. Yariv, {\it Quantum Electronics}, 2nd
Ed. (J. Wiley \& Sons, New York, 1975).

\bibitem{gauge} In Eq. (\ref{eq:gauge}), $\phi_j$ depends on the gauge
choice and the factor $A_j$ also depends on this choice.  While $A_j$
is gauge-dependent, the operator $a + a^\dag$ is not.  Since it is the
operator commutation relations, and not the specific form of $A_j$,
which are relevant, our final equations of motion are properly gauge
invariant.

\bibitem{classical} Alternatively, in this limit, we could calculate
the equations of motion from Hamilton's equations, regarding the
Hamiltonian (1) as classical, and expressing all variables as
conjugate pairs. The resulting equations of motion would then be
identical to Eqs.\ (23) - (26).

\bibitem{chakra} S. Chakravarty, G.-L. Ingold, S. Kivelson, and
A. Luther, Phys. Rev. Lett. {\bf 56}, 2303 (1986).

\bibitem{leggett} A. O. Caldeira and A. J. Leggett, Ann. Phys. (N. Y.)
{\bf 149}, 374 (1983).

\bibitem{ambeg} V. Ambegaokar, U. Eckern, and G. Sch\"{o}n,
Phys. Rev. Lett.  {\bf 48}, 1745 (1982).

\bibitem{tinkham} M. Tinkham, {\it Introduction to Superconductivity},
2nd Ed. (McGraw-Hill, New York, 1996).

\bibitem{buisson} O. Buisson and F. W. J. Hekking, cond-mat/0008275,
August 18, 2000.

\bibitem{shnirman} A. Shnirman, G. Sch{\"o}n, and Z. Hermon,
Phys. Rev. Lett. {\bf 79}, 2371 (1997).

\bibitem{asymm} The form of Eq.\ (\ref{eq:canon}) appears asymmetric,
in that the coupling between cavity and junction involves the variable
$p_r^\prime$ but not the canonically conjugate variable $q_r^\prime$.
In principle, a coupling through $q_r^\prime$ could also be present.
For such a coupling, we believe that the term $\hbar\Omega
(q_r^\prime)^2/2$ would be replaced by a term of the form
$\hbar\Omega\left(q_r^\prime + \sum_j \sqrt{2G_j} \phi_j^\prime
\right)^2/2$, where $G_j$ is the appropriate coupling strength for
this type of interaction.  This extra term would behave as a kind of
inductive coupling between the current in the cavity (represented by
the variable $q_r^\prime$) and the phase variables $\phi_j^\prime$ of
the junction.  We have chosen not to include this term in the present
work; the close resemblance between our results and experiment
suggests that this neglect is justified, though this coupling could be
significant in some experimental circumstances.

\bibitem{note2} This same picture of capacitive coupling suggests a
simple interpretation of Eq.\ (\ref{eq:eom1}).  This equation
basically expresses the voltage $\dot{\gamma}_j$ as a linear
combination of two charge variables, $n_j^\prime$ and $a_I$.  This
linear relation has the standard electrostatic form $V_i =
\sum_j(C^{-1})_{ij}Q_j$ for a system of voltages, $V_i$, linearly
related to charges, $Q_j$, by an inverse capacitance matrix, $C^{-1}$.

\bibitem{numrec} W. H. Press, S. A. Teukolsky, W. T. Vetterling, and
B. P. Flannery, {\it Numerical Recipes}, (Cambridge University Press,
NY, 1992).

\bibitem{vasilic} B. Vasilic, P. Barbara, S. V. Shitov, E. Ott, T. M.
Antonsen, and C. J. Lobb, Abstract Y27.001 of the APS March Meeting,
Seattle, WA, 2001.

\bibitem{kuramoto} Y. Kuramoto, in {\it International Symposium on
Mathematical Problems in Theoretical Physics}, edited by H. Araki,
Lecture Notes in Physics Vol. 39 (Springer, Berlin, 1975), pp. 420 --
422.

\bibitem{strogatz} See e. g., S. H. Strogatz, Physica D {\bf 143}, 1
(2000); K. Wiesenfeld, P. Colet, and S. H. Strogatz, Phys. Rev. Lett.
{\bf 76}, 404 (1996).

\bibitem{wenbin} See, for example, H. S. J. Van der Zant,
C. J. Muller, L. J.  Geerligs, C. J. P. M. Harmans, and J. E. Mooij,
Phys. Rev. B {\bf 38}, 5154 (1988); T. S. Tighe, A. T. Johnson, and
M. Tinkham, Phys. Rev. B {\bf 44}, 10286 (1991); H. S. J. Van der
Zant, F. C. Fritschy, T. P. Orlando, and J. E. Mooij, Phys.
Rev. Lett. {\bf 66}, 2531 (1991).

\bibitem{asc2000} B. Vasilic, P. Barbara, S. V. Shitov, and
C. J. Lobb, in the proceedings of Applied Superconductivity Conference
2000.

\bibitem{private} P. Barbara, private communication.

\bibitem{lobbapl} B. Vasilic, S. V. Shitov, C. J. Lobb, P. Barbara,
Appl. Phys. Lett. {\bf 78}, 1137 (2001).

\bibitem{squid} M. J. Everitt, P. Stiffel, T. D. Clark, A. Vourdas,
J. F. Ralph, H. Prance, and R. J. Prance, Phys. Rev. B {\bf 63},
144530 (2001).
\end{references}
\end{document}